\documentclass[11pt,draftcls,onecolumn]{IEEEtran}
\usepackage[cmex10]{amsmath} 
\usepackage{cite} 
\usepackage{hyperref} 
\usepackage{xcolor}
\definecolor{darkblue}{rgb}{.15,0,.7}
\hypersetup{
	colorlinks=true,
	linkcolor=darkblue,
	citecolor=darkblue,
	urlcolor=darkblue
}

\usepackage[pdftex]{graphicx} 
\usepackage{epstopdf}
\usepackage{subcaption}
\usepackage{amssymb} 
\usepackage{booktabs} 
\usepackage{tikz}
\usetikzlibrary{arrows}
\usetikzlibrary{positioning}

\usepackage{enumitem} 
\usepackage{url}
\usepackage{flushend}

\usepackage{amsthm} 
\newtheorem{theorem}{Theorem}
\newtheorem*{theorem*}{Theorem}
\newtheorem{lemma}{Lemma}
\newtheorem*{lemma*}{Lemma}
\newtheorem*{claim*}{Claim}
\newtheorem{prop}{Proposition}

\theoremstyle{definition}
\newtheorem{definition}{Definition}

\theoremstyle{remark}

\newcommand{\figurescaleparameter}{0.3}
\DeclareMathOperator*{\argmin}{arg\,min}
\DeclareMathOperator*{\arginf}{arg\,inf}
\newcommand{\bigo}{O}

\newcommand{\cdf}{cdf}
\newcommand{\datauniverse}{\mathcal{D}}
\newcommand{\databaseuniverse}{\datauniverse^n} 
\newcommand{\datauniversedim}{l}
\newcommand{\denom}{g(\epsilon)}
\newcommand{\dpmapping}{\mathcal{U}_{\epsilon}} 
\newcommand{\diff}{\Delta}
\newcommand{\estimatorset}{\hat{\mathcal{Q}}_q}
\newcommand{\expect}{\mathbb{E}} 
\newcommand{\entry}{v}
\newcommand{\gausscdf}{\Phi}
\newcommand{\ham}{d}

\newcommand{\mecrange}{\mathcal{S}} 
\newcommand{\mecmapping}{\mu} 
\newcommand{\minimax}{\mathfrak{D}_{\epsilon}}

\newcommand{\naturalnumber}{\mathbb{N}}
\newcommand{\nonnegreal}{\mathbb{R}^+} 
\newcommand{\nonnegextreal}{\overline{\mathbb{R}}^+} 
\renewcommand{\Pr}{\mathbb{P}}
\newcommand{\probmeasures}{\mathcal{P}} 
\newcommand{\powerset}[1]{\wp(#1)} 
\newcommand{\pmf}{pmf}
\newcommand{\queryrange}{\mathcal{R}} 
\newcommand{\querydis}{\rho} 
\newcommand{\Queryvar}{Z} 
\newcommand{\queryvar}{z}
\newcommand{\realnumber}{\mathbb{R}} 
\newcommand{\realspace}[1]{\mathbb{R}^{#1}}
\newcommand{\ReconX}{\widetilde{X}} 
\newcommand{\reconX}{\widetilde{x}}

\newcommand{\hamentry}{\zeta}

\begin{document}

\title{A Minimax Distortion View of Differentially Private Query Release}

\author{Weina~Wang,
		Lei~Ying,
		and Junshan~Zhang
\thanks{W. Wang, L. Ying and J. Zhang are with the School of Electrical, Computer and Energy Engineering,
Arizona State University, Tempe, AZ 85281 USA (e-mail: weina.wang@asu.edu; lei.ying.2@asu.edu; junshan.zhang@asu.edu).}}

\maketitle
\begin{abstract}
We consider the problem of differentially private query release through a synthetic database approach. Departing from the existing approaches that require the query set to be specified in advance, we advocate to devise query-set independent mechanisms, with an ambitious goal of providing accurate answers, while meeting the privacy constraints, for all queries in a general query class. Specifically, a differentially private mechanism is constructed to ``encode'' rich stochastic structure into the synthetic database, and ``customized'' companion estimators are then derived to provide accurate answers by making use of all available information, including the mechanism (which is public information) and the query functions. Accordingly, the distortion under the best of this kind of mechanisms at the worst-case query in a general query class, so called the minimax distortion, provides a fundamental characterization of differentially private query release.

For the general class of statistical queries, we prove that with the squared-error distortion measure, the minimax distortion is $\bigo(1/n)$ by deriving asymptotically tight upper and lower bounds in the regime that the database size $n$ goes to infinity. The upper bound is achievable by a mechanism $\mathcal{E}$ and its corresponding companion estimators, which points directly to the feasibility of the proposed approach in large databases. We further evaluate the mechanism $\mathcal{E}$ and the companion estimators through experiments on real datasets from Netflix and Facebook. Experimental results show improvement over the state-of-art MWEM algorithm and verify the scaling behavior $\bigo(1/n)$ of the minimax distortion.
\end{abstract}

\section{Introduction}
It is envisaged that in the forthcoming ``big data'' era, there will be an abundance of rich data about individuals in many domains, such as healthcare, mobile networks, social networks and web search. While data analysis uncovers scientific and societal insights, it also poses potential ``threats'' to personal privacy. It is therefore of great interest to establish a systematic understanding of privacy-preserving data analysis, aiming to provide utility for data analytics while preserving privacy. To rigorously quantify privacy, the celebrated notion of differential privacy, introduced in a line of work \cite{DwoMcSNis_06,Dwo_06,DwoKenMcS_06}, has emerged as an analytical foundation for privacy-preserving data analysis.

Viewing a database as a vector of rows, with each row corresponding to some sensitive record of an individual (e.g., a patient's medical record), an information releasing mechanism is said to be \emph{$\epsilon$-differentially private} if the change of a single row alters the probability of any output instance by at most an $e^{\epsilon}$ multiplicative factor. By this requirement, the presence of an individual, or the content of the record associated with an individual, cannot be exactly deduced from the released information. Therefore, a differentially private mechanism guarantees that only limited \emph{additional} information about an individual would be leaked.

As is standard, information about a database is acquired through queries. Therefore, a central problem in differential privacy is to privately release outputs that permit accurate answers to be derived for as many as possible queries. This problem has been extensively studied in the differential privacy literature, and many mechanisms for query release have been developed (see, e.g., \cite{DwoMcSNis_06,BluLigRot_08,DwoNaoRei_09,RotRou_10,DwoRotVad_10,HarRot_10,HarLigMcS_12,GupRotUll_12,GabAriHsu_14}). Adopted by much of the existing work, a natural approach is to non-interactively generate a synthetic database, which is a one-shot ``sanitization'' of the original database consisting of rows that come from the same data universe as the rows of the original database.

In contrast to the interactive counterpart, the non-interactive synthetic database approach allows arbitrary number of queries to be answered without compromising differential privacy. More specifically, queries arrive online in the interactive approach and each query consumes some privacy budget. Therefore, a privacy allocation plan is needed and only a finite number of queries can be answered before the privacy is breached. While in the non-interactive approach, the privacy budget is used all at once for the synthetic database generation, since further processing of the released synthetic database does not consume any privacy budget. As long as the synthetic database is released through a differentially private mechanism, arbitrary number of queries can be answered without compromising differential privacy.

However, although the synthetic database approach allows arbitrary number of queries to be answered without compromising differential privacy, most existing mechanisms for synthetic database release are still confined to a specific query set. Typically, the existing mechanisms  \cite{BluLigRot_08,DwoNaoRei_09,HarLigMcS_12,GupRotUll_12,GabAriHsu_14} require the query set to be specified beforehand, and the accuracy guarantee becomes worse as the size of the query set increases. There are at least two drawbacks in this approach. First, to specify a query set beforehand, a priori knowledge of the queries of interest is needed, and the queries cannot be chosen adaptively. Second, to achieve certain accuracy, the size of the query set must be smaller than a threshold. However, as pointed out in \cite{HefLig_14}, in many research settings, it is hard to decide in advance exactly which statistics should be computed. As a consequence, the synthetic database approach would not work well for such scenarios. These drawbacks debilitate the promise that arbitrary number of queries can be answered privately in a non-interactive approach, giving rise to the following question: \emph{is it possible to make the synthetic database releasing mechanism independent of any specific query set while still enabling accurate answers to be derived for all queries in a general query class from the released synthetic database?} If this could be done, the synthetic database approach would be literally ``non-interactive,'' in the sense that users do not need to interact with the curator during the entire process, whereas users need to submit the query set to the curator beforehand in the existing mechanisms.
\begin{figure*}
\centering
\begin{tikzpicture}[>=stealth',shorten >=1pt,auto,thick,text height=1.5ex,text depth=.25ex,
	main node/.style={
		rectangle,
		minimum size=0.8cm,
		minimum width=1.5cm,
		thick,
		fill=blue!20,
		draw
		},
		second node/.style={
		rectangle,
		minimum size=1cm,
		minimum width=2cm,
		thick,
		fill=white,
		draw=white
	},
	point/.style={
	coordinate,
	inner sep=0pt,
	minimum size=2pt,
	fill=red
	}]
	\matrix[row sep=5mm,column sep=5mm] {
		\node (p1) [point] {}; &
		\node (p2) [point] {}; &
		\node[main node] (mn1) {$p(y\mid x)$}; &
		\node[main node] (mn2) {$\hat q$}; &
		\node[point] {}; &
		\node[point] {}; &
		\node (p9) [point] {}; &
		\node[point] {}; &
		\node[point] {}; &
		\node[point] {};\\
		\node[point] {}; &
		\node[point] {}; &
		\node[second node,align=center] (n4) {$\epsilon$-differentially private\\ mechanism $\mathcal M$}; &
		\node[second node,align=center] (n5) {estimator\\}; &
		\node[point] {}; &
		\node[point] {}; &\node[main node] (mn4) {$\rho$}; &
		\node [point] {}; & 
		\node [point] {}; & 
		\node (p8) [point] {}; &\\
		\node[point] {}; &
		\node (p3) [point] {}; &
		\node[point] {}; &
		\node[main node] (mn3) {$q$}; &
		\node[point] {}; &
		\node[point] {}; &
		\node (p10) [point] {}; &
		\node[point] {}; &
		\node[point] {}; & 
		\node[point] {};\\
	};
	\node (p4) [point,above=of mn2,xshift=-3mm] {};
	\node (p5) [point,above=of mn2,xshift=3mm] {};
	\path (p1) edge[->] node {$x$} (mn1);
	\path (mn1) edge[->] node {$Y$} (mn2);
	\draw [->] (mn2.east) -| (mn4.north);
	\path (mn2) edge node {$\hat q(Y)$} node [swap] {released answer} (p9);
	\draw [dashed,->] (p2.south) |- (mn3.west);
	\path [dashed] (mn3) edge node {$q(x)$} node [swap] {actual answer} (p10);
	\draw [dashed,->] (mn3.east) -| (mn4.south);
	\node (p6) [point,below=of p4] {};
	\node (p7) [point,below=of p5] {};
	\path (p4) edge[->] node [swap] {$q$} (p6);
	\path (p5) edge[->] node {$p(y\mid x)$} (p7);
	\path (mn4) edge[->] node {distortion} (p8);
	\end{tikzpicture}
\caption{Road map of our approach for differentially private query release.}
\label{fig:scheme}
\end{figure*}
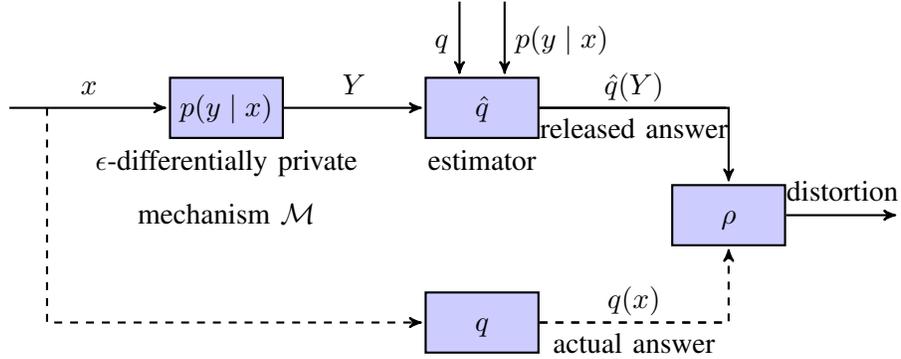

In this paper, we give positive answers to the above question for a general class of queries, via taking the following approach. First, a synthetic database is released by a query-set independent differentially private mechanism, aiming at providing accurate answers for all queries in the query class. Then each query is answered by an estimator based on the released synthetic database, rather than directly carried out as if the synthetic database were the actual database. In particular, the mechanism is constructed to ``encode'' rich stochastic structure into the synthetic database, and the estimator makes use of the structure of the mechanism (which is public information) and the query function. This approach decouples synthetic database generating and query answering. By introducing the flexibility of ``customizing'' estimators for different queries, it opens the possibility of deriving accurate answers for all queries in a general query class from the same released synthetic database. We use \emph{synthetic database release} to refer to the process of generating an output synthetic database, and \emph{query release} to refer to the entire process including releasing a synthetic database and deriving answers to queries using estimators.

Based on this approach, we advocate a minimax distortion view of differentially private query release. Consider a database consisting of $n$ rows\slash entries, each of which takes values from a domain $\datauniverse=\{0,1\}^\datauniversedim$, i.e., they have $\datauniversedim$ binary attributes.
The database is then represented by a vector $x\in\databaseuniverse$. Consider an $\epsilon$-differentially private mechanism $\mathcal{M}$ for synthetic database release and let $Y=\mathcal{M}(x)$ denote the output. For each query $q\colon\databaseuniverse\rightarrow\queryrange$ in a query class $\mathcal{Q}$, where $\queryrange$ is some abstract space, an estimator $\hat{q}\colon\databaseuniverse\rightarrow\queryrange$ is used to answer the query based on the synthetic database, and the answer is denoted by $\hat{q}(Y)$, as illustrated in Figure~\ref{fig:scheme}. The accuracy of $\mathcal{M}$ for a query $q\in\mathcal{Q}$ is evaluated when an optimal estimator $\hat{q}^*$ is in use, since an optimal estimator fully exploits the available information in the mechanism. To guarantee accuracy for all queries in the query class, the performance of $\mathcal{M}$ is measured by the worst-case distortion among queries in $\mathcal{Q}$. Then a fundamental characterization of differentially private query release is the following minimax distortion:
\begin{equation}
\minimax=\inf_{\substack{\epsilon\text{-differentially}\\\text{private mechanisms}}}\sup_{q\in\mathcal{Q},x\in\databaseuniverse}\expect[\querydis(\hat{q}^*(Y),q(x))],
\end{equation}
where $\querydis$ is a distortion measure, $\hat{q}^*$ is the optimal estimator, and $Y$ follows the probability distribution induced by $x$ through the mechanism. This minimax distortion characterizes the best one can get from an $\epsilon$-differentially private synthetic database releasing mechanism for the worst-case query accuracy guarantee, yielding a minimax distortion view of differentially private query release. Our main contributions are summarized as follows.

\subsection*{Contributions}
1) We propose a two-phase approach for differentially private query release: First, a synthetic database is released by a query-set independent differentially private mechanism, aiming at providing accurate answers for all queries in a general query class; Then queries are answered by customized estimators. Based on this approach, we advocate a minimax distortion view of differentially private query release, where the minimax distortion $\minimax$ is defined to be the distortion under the best $\epsilon$-differentially private synthetic database releasing mechanism for the worst-case query in a general query class. Accordingly, the best mechanism allows all queries in a general query class to be answered with a distortion upper bounded by the minimax distortion.

2) For the class of statistical queries (which is a generalization of the class of linear queries in the literature), we consider the minimax distortion $\minimax^{\mathrm{S}}$ with the squared-error distortion measure, i.e., $\rho(s,t)=(s-t)^2$ for any $s,t\in\realnumber$. We prove that the minimax distortion $\minimax^{\mathrm{S}}$ is $\bigo(1/n)$ by deriving asymptotically tight upper and lower bounds in the regime that the database size $n$ goes to infinity, for given data universe dimension $l$ and privacy level $\epsilon$.

The upper bound on $\minimax^{\mathrm{S}}$ is achieved by a differentially private synthetic database releasing mechanism $\mathcal{E}$ and the companion estimators. The mechanism $\mathcal{E}$ can be viewed as an instance of the exponential mechanism and the randomized response mechanism. It encodes an independence structure into the released synthetic database that is exploited by the companion estimators. Under $\mathcal{E}$ and the estimators, all the statistical queries can be answered with distortion $\bigo(1/n)$, which guarantees reasonable accuracy in large databases. In conclusion, there exists a query-set independent differentially private synthetic database releasing mechanism that permits accurate answers to be derived for all the statistical queries from the released synthetic database.

3) We evaluate the mechanism $\mathcal{E}$ and the companion estimators through a number of experiments. The experimental results on a Netflix dataset for statistical queries show that this approach provides reasonable accuracy for all the tested queries, irrespective of the form of the queries or the number of the tested queries, which improves over the MWEM algorithm. The scaling behavior $\bigo(1/n)$ of the minimax distortion is also verified by the results. The experiment on a Facebook dataset shows that this approach works well for the application of differentially private cut function release for graphs.

\subsection*{Related Work}
Differential privacy, introduced in the seminal work \cite{DwoMcSNis_06,Dwo_06}, has attracted much attention and has emerged as an analytical foundation for privacy-preserving data analysis. Extensive research has been done for both interactive and non-interactive approaches.

Non-interactive approaches have been preferred by data-mining and statistics community. However, some negative results have been found about this approach. Dinur and Nissim~\cite{DinNis_03} showed that noise of magnitude $o(\sqrt{n})$ is blatantly non-private against $n\log^2n$ random queries, where the queries may involve only a subset of the rows. Dwork et al.~\cite{DwoMcSNis_06} considered the statistical difference between two distributions that are induced by two databases that have very different answers to the same query. They showed that for many queries, this statistical difference is small unless the database size is exponential in the dimension of the data universe.

These negative results motivate interactive approaches, where the number of queries was initially limited to a sublinear order of $n$. Dwork et al.~\cite{DwoMcSNis_06} proposed the Laplace mechanism that adds Laplace noise to the real answer of a low sensitivity query. When independent noise is added to different queries, the distortion of each query scaled as $\bigo(|\mathcal{Q}|/n)$. Subsequent work \cite{RotRou_10,HarRot_10} focused on predicate/linear queries and developed mechanisms that allow exponential number of queries to be answered with distortion $\bigo(\mathrm{polylog}(|\mathcal{Q}|)/n^{1/3})$ and $\bigo((\log(|\mathcal{Q}|))^{1/2}/n^{1/2})$, respectively, where the latter is for $(\epsilon,\delta)$-differential privacy.

Non-interactive approach was revisited by Blum, Ligett and Roth \cite{BluLigRot_08}. The mechanism proposed in this work guarantees that the distortion for each predicate query in a concept class $\mathcal{Q}$ is upper bounded by $\bigo((\mathrm{VCDIM}(\mathcal{Q}))^{1/3}/n^{1/3})$, where $\mathrm{VCDIM}(\mathcal{Q})$ is the VC-dimension of $\mathcal{Q}$. A similar distortion bound $\bigo((\log(|\mathcal{Q}|))^{1/3}/n^{1/3})$ was achieved by the work of Hardt, Ligett and McSherry \cite{HarLigMcS_12} for linear queries. A distortion bound $\bigo((\log(|\mathcal{Q}|))^{1/2}/n^{1/2})$ under $(\epsilon,\delta)$-differential privacy was also achieved in this work. In this paper, we consider a more general class of queries, named statistical queries, and aim at providing accurate answers for all queries in this query class. If the absolute-error distortion $\rho=|s-t|$ for any $s,t\in\realnumber$ is used, as the above related work, then the synthetic database releasing mechanism $\mathcal{E}$ and the proposed companion estimators give answers to all the statistical queries with expected distortion $\bigo(1/n^{1/2})$.

Minimax risk is a classical framework in statistics \cite{Tsy_09} that focuses on estimating parameters of the underlying distribution. Minimax rates were studied under \emph{local privacy}, which is a privacy notion different from differential privacy, by Duchi, Jordan and Wainwright \cite{DucJorWai_13,DucJorWai_13_1}. In contrast, this study does not assume any knowledge of the underlying distribution of the database, and focuses on providing accurate answers to a general class of queries.

\subsection*{Paper Organization}
The rest of the paper is organized as follows. In Section~\ref{sec:model}, we describe the model used in this paper. In Section~\ref{sec:minimax}, we present our minimax distortion view of the differentially private query release. The class of statistical queries is studied in Section~\ref{sec:statistical}, and some generalizations are given in Section~\ref{sec:general}. Experimental evaluation of the proposed approach and the application of cut function release for graphs are presented in Section~\ref{sec:experiment}. Finally, we conclude our work and discuss future work in Section~\ref{sec:conclusion}.

\subsubsection*{Notation}
Throughout this paper we use the following basic notation. Denote the set of real numbers by $\realnumber$, the set of nonnegative real numbers by $\nonnegreal$. Let $\nonnegextreal=\nonnegreal\cup\{+\infty\}$. Denote the set of nonnegative integers by $\naturalnumber$ and denote $[n]=\{1,2,\dots,n\}$ for $n\in\naturalnumber\setminus\{0\}$.

\section{Model}\label{sec:model}
We consider the following model for a database. A database is represented by a vector $x$ of length $n$, with each entry corresponding to a row of the database and $n$ being the size of the database. Entries of $x$ are denoted by $x_1,x_2,\dots,x_n$, and they take values from a domain $\datauniverse=\{0,1\}^\datauniversedim$, i.e., they have $l$ binary attributes.
Then $\databaseuniverse=(\{0,1\}^\datauniversedim)^n$ denotes the set of all possible databases. Two databases $x,x'\in\databaseuniverse$ are said to be \emph{neighbors} if they differ on exactly one row, and $x\sim x'$ denotes the neighboring relation.

Information about a database is acquired through queries. A \emph{query} is a function $q\colon \databaseuniverse\rightarrow\queryrange$, where $\mathcal{R}$ is some abstract range. Consider a database $x\in\databaseuniverse$. The answer $q(x)$ to the query contains information about $x$; however, directly releasing $q(x)$ may compromise privacy, necessitating privacy-preserving information releasing mechanisms.

\begin{definition}
A \emph{mechanism} $\mathcal{M}$ is specified by an \emph{associated mapping} $\mecmapping_{\mathcal{M}}\colon \databaseuniverse\rightarrow \probmeasures$, where $\probmeasures$ is the set of probability measures on some measurable space $(\mecrange,\mathcal{F})$, called the \emph{range} of the mechanism $\mathcal{M}$. Taking a database $x\in\databaseuniverse$ as the input, the mechanism $\mathcal{M}$ outputs an $\mecrange$-valued random variable with distribution measure $\mecmapping_{\mathcal{M}}(x)$ on $(\mecrange,\mathcal{F})$.
\end{definition}

\begin{definition}
(Dwork et al. \cite{DwoMcSNis_06,Dwo_06})
A mechanism $\mathcal{M}$ is $\epsilon$\emph{-differentially private} for some $\epsilon\in\overline{\mathbb{R}}^+$ if for any pair of neighboring databases $x,x'\in\databaseuniverse$, and any measurable $\mathcal{K}\in\mathcal{F}$,
\begin{equation}\label{eqDP}
\Pr\{\mathcal{M}(x)\in\mathcal{K}\}\le e^\epsilon\Pr\{\mathcal{M}(x')\in \mathcal{K}\}.
\end{equation}
\end{definition}

Intuitively, differential privacy requires certain indistinguishability between the distributions induced by neighboring databases. The smaller $\epsilon$ is, the more indistinguishability is required, and hence the better privacy is. We call the parameter $\epsilon$ the \emph{level of differential privacy}. Note that the differential privacy property of a mechanism is fully characterized by its associated mapping.

We consider differentially private mechanisms for non-interactive synthetic database release. Specifically, let $\powerset{\databaseuniverse}$ denote the power set of $\databaseuniverse$. Then we consider differentially private mechanisms with range $(\databaseuniverse,\powerset{\databaseuniverse})$. Let $\mathcal{M}$ be such a mechanism and $x\in\databaseuniverse$ be a database. Then the output $Y=\mathcal{M}(x)$ is a $\databaseuniverse$-valued random variable that represents the released synthetic database. Many mechanisms for synthetic database release have been developed (see, e.g., \cite{BluLigRot_08,DwoNaoRei_09,HarLigMcS_12,GupRotUll_12,GabAriHsu_14}), where a query $q$ is typically answered by $q(Y)$, i.e., a query is answered as if the synthetic database were the actual database. These mechanisms require the query set to be specified in advance and the accuracy guarantee depends on the size of the query set.

In this paper, we explore the following approach. First, a synthetic database is released using a query-set independent differentially private mechanism, and then queries are answered by customized estimators. For each query $q$ in a query class $\mathcal{Q}$, an estimator $\hat{q}\colon\databaseuniverse\rightarrow\queryrange$ is used to answer the query based on the synthetic database, and thus the answer is denoted by $\hat{q}(Y)$. To achieve good accuracy, the estimator $\hat{q}$ should be designed according to the mechanism $\mathcal{M}$ and the query $q$, making use of all the available information. The distortion between the actual answer $q(x)$ and the released answer $\hat{q}(Y)$ is measured by a distortion measure $\querydis$ on the range of the query $q$. This approach is illustrated in Figure~\ref{fig:scheme}, where the mechanism is represented by the probability distribution $p(\cdot\mid x)$ of $Y$ for each input database $x$, and $\hat{q}$ has $p(\cdot\mid x)$ and $q$ as inputs to indicate the design dependence.

Note that in this non-interactive approach, as long as the mechanism $\mathcal{M}$ is $\epsilon$-differentially private, the whole query release process is $\epsilon$-differentially private, i.e., arbitrary number of queries can be answered and any estimator can be used, with the level of differential privacy still preserved.

\section{Minimax Distortion}\label{sec:minimax}
The proposed approach aims at privately releasing a synthetic database that permits accurate answers to be derived for all queries in a query class. Therefore, a natural fundamental characterization of differentially private query release is the following minimax distortion: the distortion under the best differentially private synthetic database releasing mechanism (the ``min'' part) for the worst-case query in the query class (the ``max'' part).

For each query $q\colon\databaseuniverse\rightarrow\queryrange$, let $\querydis\colon \queryrange\times\queryrange\rightarrow\nonnegreal$ be a distortion measure on the space $\mathcal{R}$. For the sake of fair comparison, we assume that $q$ is normalized, i.e.,
\begin{equation}
\max_{x,x'\in\databaseuniverse}\querydis(q(x),q(x'))= 1,
\end{equation}
which rules out trivial queries that map all possible databases to a constant. For each query $q$, to guarantee that the released answers have ``physical meanings,'' we consider the estimators such that the answers released by them correspond to possible answers to the query $q$ on real databases, i.e., the estimators in $\estimatorset=\{\hat{q}\rightarrow\queryrange\mid\hat{q}(\databaseuniverse)\subseteq q(\databaseuniverse)\}$, which we call \emph{proper estimators}. Consider an $\epsilon$-differentially private mechanism $\mathcal{M}$ and an estimator $\hat{q}\in\estimatorset$ for the query $q$, the distortion of the answer is defined by the following worst-case distortion among all possible databases:
\begin{equation}
\sup_{x\in\databaseuniverse}\expect_{Y\sim\mecmapping_{\mathcal{M}}(x)}[\querydis(\hat{q}(Y),q(x))],
\end{equation}
where the subscript $Y\sim\mecmapping_{\mathcal{M}}(x)$ indicates that $Y$ follows the distribution $\mecmapping_{\mathcal{M}}(x)$, and the expectation is taken over all the randomness.

To minimize distortion, an estimator should be designed according to the mechanism $\mathcal{M}$ and the query $q$, making use of all the available information. Therefore an optimal estimator $\hat{q}^*$ is given by
\begin{equation}\label{eq:opt_est}
\hat{q}^*\in\arginf_{\hat{q}\in\estimatorset}\sup_{x\in\databaseuniverse}\expect_{Y\sim\mecmapping_{\mathcal{M}}(x)}[\querydis(\hat{q}(Y),q(x))].
\end{equation}
Note that the set $\estimatorset$ contains only a finite number of estimators since it consists of mappings from $\databaseuniverse$ to $q(\databaseuniverse)$, which are both finite sets, indicating that the infimum in \eqref{eq:opt_est} can be attained. Since the information in a mechanism is fully exploited only when an optimal estimator is in use, the accuracy of an $\epsilon$-differentially private mechanism $\mathcal{M}$ for a query $q$ is evaluated with an optimal estimator $\hat{q}^*$, i.e., by the distortion
\begin{equation}
\sup_{x\in\databaseuniverse}\expect_{Y\sim\mecmapping_{\mathcal{M}}(x)}[\querydis(\hat{q}^*(Y),q(x))].
\end{equation}

The synthetic database released by $\mathcal{M}$ is expected to answer all queries in a query class $\mathcal{Q}$. To guarantee accuracy for all queries in $\mathcal{Q}$, the performance of $\mathcal{M}$ is measured by the worst-case distortion among all queries in $\mathcal{Q}$, i.e., by
\begin{equation}
\sup_{q\in\mathcal{Q}}\biggl\{\sup_{x\in\databaseuniverse}\expect_{Y\sim\mecmapping_{\mathcal{M}}(x)}[\querydis(\hat{q}^*(Y),q(x))]\biggr\}.
\end{equation}
Let $\dpmapping$ be the set of mappings associated with $\epsilon$-differentially private mechanisms. Then we define the \emph{minimax distortion} as
\begin{equation}\label{eq:minimax}
\minimax=\inf_{\mecmapping_{\mathcal{M}}\in\dpmapping}\sup_{q\in\mathcal{Q}}\biggl\{\sup_{x\in\databaseuniverse}\expect_{Y\sim\mecmapping_{\mathcal{M}}(x)}[\querydis(\hat{q}^*(Y),q(x))]\biggr\}.
\end{equation}
The minimax distortion is a fundamental characterization of $\epsilon$-differentially private query release since it characterizes the best one can get from an $\epsilon$-differentially private synthetic database releasing mechanism for the worst-case query accuracy guarantee. In what follows we will study differentially private query release from this minimax distortion view, and derive upper and lower bounds on the minimax distortion accordingly.

\section{Statistical Queries}\label{sec:statistical}
In this section, we consider differentially private query release for the class of statistical queries, which is a much larger class than the class of linear queries in the literature.

\begin{definition}\label{def:stat_query}
A \emph{statistical query} $q_{\varphi}\colon\databaseuniverse\rightarrow\mathbb{R}$ is specified by a sequence of functions
\begin{equation}
\varphi=(\varphi_i\colon \datauniverse\rightarrow\mathbb{R},i=1,2,\dots),
\end{equation}
where each $\varphi_i$ is a function of the $i$th row of the database, which we call a \emph{row function}, and there is no constraint on its form except boundedness. Let $a_i=\min_{\entry\in\datauniverse}\varphi_i(\entry)$, $b_i=\max_{\entry\in\datauniverse}\varphi_i(\entry)$ and $c_i=b_i-a_i$. Assume that for any $i\in[n]$, $a\le a_i< b_i\le b$ and $c_i\ge c$ for some $a,b,c\in\mathbb{R}$ with $c>0$. Then $q_{\varphi}$ is defined by
\begin{equation}\label{eq:stat}
q_{\varphi}(x)=\frac{1}{\sum_{i=1}^{n}c_i}\sum_{i=1}^{n}\varphi_i(x_i),
\end{equation}
where $x_1,\dots,x_n$ are the rows of the database $x$.
\end{definition}

Note that the above definition of statistical query is a generalization of the so called \emph{linear query} (and its special form \emph{predicate}\slash\emph{counting query}) in the literature \cite{BluLigRot_08,RotRou_10,HarTal_10,HarRot_10,LiHayRas_10,HarLigMcS_12,GupRotUll_12,Ull_13}, since a linear query can be written as a statistical query with identical row functions for all the rows. Linear queries can be answered as long as the histogram of a database is known. However, histograms are often not sufficient for answering statistical queries, making the approaches that privately release histograms not applicable for statistical queries.

Denote the class of statistical queries by $\mathcal{Q}^{\mathrm{S}}$ and let $\querydis\colon\realnumber\times\realnumber\rightarrow\nonnegreal$ be the squared-error distortion, i.e., $\querydis(s,t)=(s-t)^2$ for any $s,t\in\realnumber$. Then the minimax distortion for statistical queries can be written as
\begin{equation}
\minimax^{\mathrm{S}}=\inf_{\mecmapping_{\mathcal{M}}\in\dpmapping}\sup_{q_{\varphi}\in\mathcal{Q}^{\mathrm{S}},x\in\databaseuniverse}\expect_{Y\sim\mecmapping_{\mathcal{M}}(x)}\bigl[|\hat{q}_{\varphi}^*(Y)-q_{\varphi}(x)|^2\bigr].
\end{equation}
\begin{theorem}\label{thm:mini_statistical}
The minimax distortion for statistical queries satisfies the following bounds:
\begin{equation}
\frac{\bigl(1-\gausscdf(1)\bigr)^2}{ 2^{l+4}\bigl(1+\frac{e^{\epsilon}}{2^l-1}\bigr)^3}\frac{1}{n}+o\biggl(\frac{1}{n}\biggr)\le \minimax^{\mathrm{S}}\le\frac{4(b-a)^2\bigl(1+(2^l-1)e^{-\epsilon}\bigr)^2}{c^2(1-e^{-\epsilon})^2}\frac{1}{n},
\end{equation}
where $\gausscdf$ is the cumulative distribution function (\cdf) of the standard Gaussian distribution, and $a,b,c$ are the constants in Definition~\ref{def:stat_query}.
\end{theorem}

The upper bound in this theorem is given by the performance of an $\epsilon$-differentially private synthetic database releasing mechanism $\mathcal{E}$ and the companion estimators, which are presented in Section~\ref{subsec:upper}. The lower bound in this theorem is derived by bounding the average distortion. The minimax distortion is defined for the worst-case distortion over statistical queries and databases. We consider a stochastic model for the queries and the database. Then the average distortion under this model serves as a lower bound on the worst-case distortion. Analyzing the average distortion under the constraint of $\epsilon$-differential privacy as in Section~\ref{subsec:lower} gives the lower bound.

Consider the asymptotic regime that the database size $n$ goes to infinity for given data universe dimension $l$ and privacy level $\epsilon$. Then the upper bound indicates that there exist query-set independent differentially private synthetic database releasing mechanisms and estimators such that all the statistical queries can be answered with distortion $\bigo(1/n)$. Further, the lower bound and the upper bound are of the same order in terms of database size, which shows that these bounds are asymptotically tight in the considered regime. We derive these bounds in the following subsections. 

\emph{Remark.} We caution that when the privacy level $\epsilon$ also scales, the upper and lower bounds given here may not meet. For example, let $\epsilon=n^{-\beta}$ for some $\beta>0$ and consider the joint asymptotic regime on the 2-dimensional $(n,\frac{1}{\epsilon})$-plane. In this case, the upper and lower bounds differ by a factor of the order of $n^{2\beta}$.

\subsection{Upper Bound on the Minimax Distortion}\label{subsec:upper}
In this subsection, we consider a specific $\epsilon$-differentially private mechanism $\mathcal{E}$ and develop the estimators companioned with it for statistical queries. Since the minimax distortion for statistical queries can be written as
\begin{equation}\label{eq:fullMinimax}
\minimax^{\mathrm{S}}=\inf_{\mecmapping_{\mathcal{M}}\in\dpmapping}\sup_{q_{\varphi}\in\mathcal{Q}^{\mathrm{S}}}\biggl\{\inf_{\hat{q}_{\varphi}\in\hat{\mathcal{Q}}_{q_{\varphi}}}\sup_{x\in\databaseuniverse}\expect_{Y\sim\mecmapping_{\mathcal{M}}(x)}\bigl[|\hat{q}_{\varphi}(Y)-q_{\varphi}(x)|^2\bigr]\biggr\},
\end{equation}
the distortion under the mechanism $\mathcal{E}$ and the developed estimators serves as an upper bound on $\minimax^{\mathrm{S}}$, which proves the upper bound in Theorem~\ref{thm:mini_statistical}. Since we only consider the mechanism $\mathcal{E}$ in this subsection, we drop the subscript $Y\sim\mecmapping_{\mathcal{E}}(x)$ from expectations for conciseness.

Consider a synthetic database releasing mechanism $\mathcal{E}$ with associated mapping $\mecmapping_{\mathcal{E}}$. For each database $x\in\databaseuniverse$, since the output $\mathcal{E}(x)$ has a discrete alphabet $\databaseuniverse$, we use the \pmf\ $p_{\mathcal{E}(x)}$ to represent the distribution measure $\mecmapping_{\mathcal{E}}(x)$. Then let the mechanism $\mathcal{E}$ be specified by
\begin{equation}\label{eq:expMec}
p_{\mathcal{E}(x)}(y)=\frac{e^{-\epsilon \ham(x,y)}}{\bigl(1+(2^l-1)e^{-\epsilon}\bigr)^n},\quad x,y\in\databaseuniverse,
\end{equation}
where $\epsilon\in\nonnegreal$ and $\ham$ is the Hamming distance on $\databaseuniverse$. By the form of $p_{\mathcal{E}(x)}$, this mechanism can be cast as an instance of the exponential mechanism with score function $-d$ \cite{McSTal_07}.

Let $Y$ denote $\mathcal{E}(x)$ for conciseness when it is clear from the context that $x$ is the underlying database. Then the \pmf\ $p_{\mathcal{E}(x)}$ can be written as
\begin{equation}\label{eq:expMecProd}
p_{Y}(y)=\prod_{i=1}^{n}\frac{e^{-\epsilon \delta(x_i,y_i)}}{1+(2^l-1)e^{-\epsilon}},\quad y\in\databaseuniverse,
\end{equation}
where $\delta(x_i,y_i)=0$ if $x_i=y_i$ and $\delta(x_i,y_i)=1$ otherwise. Let $Y_i$ denote the $i$th row of $Y$. Due to the product form above, the entries $\{Y_i,i\in[n]\}$ are independent and each entry $Y_i$ has the following \pmf\
\begin{equation}\label{eq:pmf_row}
p_{Y_i}(y_i)=\frac{e^{-\epsilon\delta(x_i,y_i)}}{1+(2^l-1)e^{-\epsilon}},\quad y_i\in\datauniverse.
\end{equation}
Therefore this mechanism can also be viewed as a randomized response scheme, where each individual's data is perturbed independently and then the perturbed data is released. Note that the mechanism $\mathcal{E}$ can be implemented distributedly due to the independence.

The differential privacy property of the mechanism $\mathcal{E}$ is given in the following lemma. The proof is standard and thus we omit it here due to space limit.
\begin{lemma}\label{lem:mecE_dp}
The mechanism $\mathcal{E}$ is $\epsilon$-differentially private.
\end{lemma}

Next we present the estimators companioned with the mechanism $\mathcal{E}$ for the class of statistical queries. Let $\denom=1+(2^l-1)e^{-\epsilon}$. For each $q_{\varphi}\in\mathcal{Q}^{\mathrm{S}}$, consider the estimator $\hat{q}_{\varphi}^{\mathrm{u}}\colon\databaseuniverse\rightarrow\queryrange$ defined by
\begin{equation}\label{eq:qhatu}
\hat{q}_{\varphi}^{\mathrm{u}}(y)=\frac{\denom}{1-e^{-\epsilon}}q_{\varphi}(y)-\frac{e^{-\epsilon}}{1-e^{-\epsilon}}C_{\varphi},
\end{equation}
where
\begin{equation}
C_{\varphi}=\frac{1}{\sum_{i=1}^{n}c_i}\sum_{i=1}^{n}\sum_{\entry\in\datauniverse}\varphi_i(\entry).
\end{equation}
\begin{lemma}\label{lem:qhatu}
Under the mechanism $\mathcal{E}$, the estimator $\hat{q}_{\varphi}^{\mathrm{u}}$ is unbiased, i.e., for any database $x\in\databaseuniverse$,
\begin{equation}
\expect[\hat{q}_{\varphi}^{\mathrm{u}}(Y)]=q(x),
\end{equation}
and the distortion of $\hat{q}_{\varphi}^{\mathrm{u}}$ satisfies the following upper bound:
\begin{equation}\label{eq:qhatu_dis}
\sup_{x\in\databaseuniverse}\expect\bigl[|\hat{q}_{\varphi}^{\mathrm{u}}(Y)-q_{\varphi}(x)|^2\bigr]\le\frac{(b-a)^2\bigl(1+(2^l-1)e^{-\epsilon}\bigr)^2}{c^2(1-e^{-\epsilon})^2}\frac{1}{n},
\end{equation}
where $a,b,c$ are the constants in Definition~\ref{def:stat_query}.
\end{lemma}

The proof of this lemma is given in Appendix~\ref{app:lem_qhatu}. The intuition is that the mechanism $\mathcal{E}$ perturbs each row of the underlying database independently, which encodes an independence structure into the released synthetic base, and then the estimator $\hat{q}_{\varphi}^{\mathrm{u}}$ exploits this structure. By the law of large numbers (LLN), the aggregate perturbation converges to the expectation, which is a constant determined by the query and thus can be removed in the estimator.

Next we present a proper estimator designed based on $\hat{q}_{\varphi}^{\mathrm{u}}$. The answer given by the estimator $\hat{q}_{\varphi}^{\mathrm{u}}$ may not always be consistent with an actual database, in which case $\hat{q}_{\varphi}^{\mathrm{u}}\notin\hat{\mathcal{Q}}_{q_{\varphi}}$. Thus we consider the estimator $\hat{q}_{\varphi}\colon\databaseuniverse\rightarrow\queryrange$ defined by
\begin{equation}\label{eq:qhat}
\hat{q}_{\varphi}(y)\in\argmin_{r\in q_{\varphi}(\databaseuniverse)}|\hat{q}_{\varphi}^{\mathrm{u}}(y)-r|,
\end{equation}
which quantizes the answer given by $\hat{q}_{\varphi}^{\mathrm{u}}$ to the closest value in $q_{\varphi}(\databaseuniverse)$. This quantization guarantees that $\hat{q}_{\varphi}$ is a proper estimator, and degrades the performance guarantee only by a factor of $4$ as shown in the following lemma, the proof of which is given in Appendix~\ref{app:lem_qhat}.
\begin{lemma}\label{lem:qhat_distortion}
Under the mechanism $\mathcal{E}$, the distortion of the estimator $\hat{q}_{\varphi}$ satisfies the following upper bound:
\begin{equation}\label{eq:qhat_dis}
\sup_{x\in\databaseuniverse}\expect\bigl[|\hat{q}_{\varphi}(Y)-q_{\varphi}(x)|^2\bigr]\le\frac{4(b-a)^2\bigl(1+(2^l-1)e^{-\epsilon}\bigr)^2}{c^2(1-e^{-\epsilon})^2}\frac{1}{n},
\end{equation}
where $a,b,c$ are the constants in Definition~\ref{def:stat_query}.
\end{lemma}

Consider the asymptotic regime that the database size $n$ goes to infinity for given data universe dimension $l$ and privacy level $\epsilon$. By the upper bounds \eqref{eq:qhatu_dis} and \eqref{eq:qhat_dis}, the estimators $\hat{q}_{\varphi}^{\mathrm{u}}$ and $\hat{q}_{\varphi}$ answer all the statistical queries with distortion $\bigo(1/n)$ based on the synthetic database released by the mechanism $\mathcal{E}$. Therefore all the statistical queries can be answered with reasonable accuracy guarantee in large databases.

Compared with existing approaches, the synthetic database releasing mechanism $\mathcal{E}$ does not require a priori knowledge of the queries of interest, and instead of answering query $q_{\varphi}$ by $q_{\varphi}(Y)$, the estimators $\hat{q}_{\varphi}^{\mathrm{u}}$ and $\hat{q}_{\varphi}$ make more use of the stochastic structure in $Y$ encoded by the mechanism $\mathcal{E}$.

\emph{Remark.} Under the absolute-error distortion defined by $\rho(s,t)=|s-t|$, for any $s,t\in\realnumber$, the distortion upper bounds for the estimators $\hat{q}_{\varphi}^{\mathrm{u}}$ and $\hat{q}_{\varphi}$ become
\begin{equation*}\label{eq:qhatu_dis_abs}
\sup_{x\in\databaseuniverse}\expect\bigl[|\hat{q}_{\varphi}^{\mathrm{u}}(Y)-q_{\varphi}(x)|\bigr]\le\frac{(b-a)\bigl(1+(2^l-1)e^{-\epsilon}\bigr)}{c(1-e^{-\epsilon})}\frac{1}{\sqrt{n}}
\end{equation*}
\begin{equation*}\label{eq:qhat_dis_abs}
\sup_{x\in\databaseuniverse}\expect\bigl[|\hat{q}_{\varphi}(Y)-q_{\varphi}(x)|\bigr]\le \frac{2(b-a)\bigl(1+(2^l-1)e^{-\epsilon}\bigr)}{c(1-e^{-\epsilon})}\frac{1}{\sqrt{n}}
\end{equation*}
since by Jensen's inequality $\bigl(\expect\bigl[|X|\bigr]\bigr)^2\le\expect\bigl[|X|^2\bigr]$ for any random variable $X$.

\emph{Remark.} By the form of the estimator $\hat{q}_{\varphi}^{\mathrm{u}}$ in \eqref{eq:qhatu}, the value $C_{\varphi}=\frac{1}{\sum_{i=1}^{n}c_i}\sum_{i=1}^{n}\sum_{\entry\in\datauniverse}\varphi_i(\entry)$ is needed to answer the query $q_{\varphi}$. In many cases, this value can be easily obtained rather than exhaustive calculation. In such case, the computation in $\hat{q}_{\varphi}^{\mathrm{u}}$ is very efficient. Take the following predicate query for an example. Recall that any $\entry\in\datauniverse=\{0,1\}^l$ is a binary vector $v=(v_1,\dots,v_l)$ of length $l$. Consider the predicate function $s(\entry)=\entry_{j_1}\cdot\entry_{j_2}\cdot\dots\entry_{j_k}$ for some $\{j_1,\dots,j_k\}$ with $1\le k\le l$, which counts the fraction of rows in the database that have value $1$ for attributes $j_1,\dots,j_k$. This predicate query is a statistical query $q_{\varphi}$ with $\varphi_i=s$ for any $i\in[n]$. The value $C_{\varphi}$ for this query is $C_{\varphi}=2^{l-k}$, which can be obtained by simple analysis.

\emph{Remark.} The estimator $\hat{q}_{\varphi}^{\mathrm{u}}$ is more computationally efficient than the estimator $\hat{q}_{\varphi}$ since it does not need to find the value closest to $\hat{q}_{\varphi}^{\mathrm{u}}(Y)$ in $q_{\varphi}(\databaseuniverse)$. Therefore when we are not constricted to the estimators in $\hat{\mathcal{Q}}_{q_{\varphi}}$, it is more desirable to use the estimator $\hat{q}_{\varphi}^{\mathrm{u}}$ from an implementation perspective.

\subsection{Lower Bound on the Minimax Distortion}\label{subsec:lower}
Consider any $\epsilon$-differentially private mechanism $\mathcal{M}$. For any query $q_{\varphi}\in\mathcal{Q}^{\mathrm{S}}$, the form of the optimal estimator depends on $q_{\varphi}$. Therefore with slight abuse of notation, we denote the optimal estimator by the function $\hat{q}^*\colon \databaseuniverse\times \mathcal{Q}^{\mathrm{S}}\rightarrow \realnumber$ and the answer by $\hat{q}^*(Y,q_{\varphi})$, where $Y$ is the synthetic database released by the mechanism $\mathcal{M}$. Then our goal is to derive a lower bound on the following worst-case distortion:
\begin{equation}
\sup_{q_{\varphi}\in\mathcal{Q}^{\mathrm{S}},x\in\databaseuniverse}\expect_{Y\sim\mecmapping_{\mathcal{M}}(x)}\bigl[|\hat{q}^*(Y,q_{\varphi})-q_{\varphi}(x)|^2\bigr].
\end{equation}

Consider such a type of queries, each of which is specified by an element $\queryvar\in\databaseuniverse$ and defined by
\begin{equation*}
q_{\queryvar}(x)=\frac{1}{n}d(x,\queryvar),\quad x\in\databaseuniverse,
\end{equation*}
where $d$ is the Hamming distance on $\databaseuniverse$. For any $\entry,\entry'\in\datauniverse$, let $\delta(\entry,\entry')=0$ if $\entry=\entry'$ and $\delta(\entry,\entry')=1$ otherwise. Then the query $q_{\queryvar}$ can be written as
\begin{equation*}
q_{\queryvar}(x)=\frac{1}{n}\sum_{i=1}^{n}\delta(x_i,\queryvar_i),
\end{equation*}
from which we can see that the query $q_{\queryvar}$ is a statistical query. Let
\begin{equation}
\mathcal{Q}^{\mathrm{\Queryvar}}=\biggl\{q_{\queryvar}\colon \databaseuniverse\rightarrow\realnumber\biggm| q_{\queryvar}(x)=\frac{1}{n}d(x,\queryvar),\queryvar\in\databaseuniverse\biggr\}.
\end{equation}
Then $\mathcal{Q}^{\mathrm{\Queryvar}}\subseteq\mathcal{Q}^{\mathrm{S}}$, and therefore
\begin{equation*}
\begin{split}
&\mspace{18mu}\sup_{q_{\varphi}\in\mathcal{Q}^{\mathrm{S}},x\in\databaseuniverse}\expect_{Y\sim\mecmapping_{\mathcal{M}}(x)}\bigl[|\hat{q}^*(Y,q_{\varphi})-q_{\varphi}(x)|^2\bigr]\\
&\ge\sup_{q_{\queryvar}\in\mathcal{Q}^{\mathrm{\Queryvar}},x\in\databaseuniverse}\expect_{Y\sim\mecmapping_{\mathcal{M}}(x)}\bigl[|\hat{q}^*(Y,q_{\queryvar})-q_{\queryvar}(x)|^2\bigr].
\end{split}
\end{equation*}

To derive a lower bound on the above supremum, consider $\databaseuniverse$-valued random variables $X,Y,\Queryvar$ with the following distributions. The random variable $X$ follows a uniform distribution, i.e., the probability mass function (\pmf) $p_X(x)=\frac{1}{2^{nl}}$ for any $x\in\databaseuniverse$. Given $X=x$, the conditional \pmf\ of $Y$ is specified by the distribution measure $\mecmapping_{\mathcal{M}}(x)$, i.e., $p_{Y\mid X}(y\mid x)=\Pr\{\mathcal{M}(x)=y\}$ for any $y\in\databaseuniverse$. The random variable $\Queryvar$ is independent of $X$ and $Y$, and it also follows a uniform distribution, i.e., the \pmf\ $p_{\Queryvar}(\queryvar)=\frac{1}{2^{nl}}$ for any $\queryvar\in\databaseuniverse$.

Consider the query $q_{\Queryvar}$, which is the query in $\mathcal{Q}^{\mathrm{Z}}$ specified by $\Queryvar$. Then $q_{\Queryvar}$ is a query chosen from $\mathcal{Q}^{\mathrm{\Queryvar}}$ uniformly at random. Due to the independence between $\Queryvar$ and $(X,Y)$, given any $X=x$ and $\Queryvar=\queryvar$, the conditional \pmf\ $p_{Y\mid X,\Queryvar}(y\mid x,\queryvar)=p_{Y\mid X}(y\mid x)$, which corresponds to $\mecmapping_{\mathcal{M}}(x)$. Therefore
\begin{align*}
&\mspace{21mu}\sup_{q_{\queryvar}\in\mathcal{Q}^{\mathrm{\Queryvar}},x\in\databaseuniverse}\expect_{Y\sim\mecmapping_{\mathcal{M}}(x)}\bigl[|\hat{q}^*(Y,q_{\queryvar})-q_{\queryvar}(x)|^2\bigr]\nonumber\\
&=\sup_{q_{\queryvar}\in\mathcal{Q}^{\mathrm{\Queryvar}},x\in\databaseuniverse}\expect\bigl[|\hat{q}^*(Y,q_{\Queryvar})-q_{\Queryvar}(X)|^2\bigm|X=x,\Queryvar=\queryvar\bigr]\\
&\ge\mspace{-20mu} \sum_{\queryvar\in\databaseuniverse,x\in\databaseuniverse}\mspace{-20mu}\expect\bigl[|\hat{q}^*(Y,q_{\Queryvar})-q_{\Queryvar}(X)|^2\bigm|X=x,\Queryvar=\queryvar\bigr]p_X(x)p_{\Queryvar}(\queryvar)\nonumber\\
&=\expect\bigl[|\hat{q}^*(Y,q_{\Queryvar})-q_{\Queryvar}(X)|^2\bigr].
\end{align*}
Note that we construct the random variables $X$ and $\Queryvar$ only for the proof. Our result in Theorem~\ref{thm:mini_statistical} does not assume any stochastic model for the database or the query. Note that $\hat{q}^*(Y,q_{\Queryvar})$ is a function of $Y$ and $\Queryvar$. Since the conditional expectation is precisely the minimum mean square estimator~\cite{AthLah_06}, we have
\begin{align}
&\mspace{24mu}\expect\bigl[|\hat{q}^*(Y,q_{\Queryvar})-q_{\Queryvar}(X)|^2\bigr]\nonumber\\
&\ge\expect\bigl[|\expect[q_{\Queryvar}(X)\mid Y,\Queryvar]-q_{\Queryvar}(X)|^2\bigr]\\
&=\frac{1}{n^2}\expect\bigl[|\expect[d(X,\Queryvar)\mid Y,\Queryvar]-d(X,\Queryvar)|^2\bigr].\label{eq:conv-ham}
\end{align}

Recall that the conditional \pmf\ $p_{Y\mid X}(\cdot\mid x)$ is specified by the distribution measure $\mecmapping_{\mathcal{M}}(x)$. Then since the mechanism $\mathcal{M}$ is $\epsilon$-differentially private, for any neighboring $x,x'\in\databaseuniverse$ and any $y\in\databaseuniverse$,
\begin{equation*}
p_{Y\mid X}(y\mid x)\le e^{\epsilon}p_{Y\mid X}(y\mid x').
\end{equation*}
This inequality is needed in the proof of the following lemma, which gives a lower bound on the expectation in \eqref{eq:conv-ham}.
\begin{lemma}\label{lem:lower}
There exists a constant $C$ such that
\begin{equation}
\begin{split}
&\mspace{24mu}\expect\bigl[|\expect[d(X,\Queryvar)\mid Y,\Queryvar]-d(X,\Queryvar)|^2\bigr]\\
&\ge\frac{1}{4}\biggl(\bigl(1-\gausscdf(1)\bigr)\sigma\gamma^{\frac{3}{2}}\sqrt{n}-\frac{C\rho\gamma}{\sigma^3}\biggr)^2,
\end{split}
\end{equation}
where $\gausscdf$ is the \cdf\ of the standard Gaussian distribution,
\begin{equation}
\gamma=\frac{1}{2\bigl(1+\frac{e^{\epsilon}}{2^l-1}\bigr)},\quad\sigma^2=\frac{1}{2^{l-1}},\quad\rho=\frac{1}{2^{l-1}}.
\end{equation}
\end{lemma}
The proof is presented in Appendix~\ref{app:lem_lower}. By this lemma, for any $\epsilon$-differentially private mechanism $\mathcal{M}$, the distortion is lower bounded as
\begin{equation}
\begin{split}
&\mspace{24mu}\sup_{q_{\varphi}\in\mathcal{Q}^{\mathrm{S}},x\in\databaseuniverse}\expect_{Y\sim\mecmapping_{\mathcal{M}}(x)}\bigl[|\hat{q}^*(Y,q_{\varphi})-q_{\varphi}(x)|^2\bigr]\\
&\ge\frac{\bigl(1-\gausscdf(1)\bigr)^2}{ 2^{l+4}\bigl(1+\frac{e^{\epsilon}}{2^l-1}\bigr)^3}\frac{1}{n}+o\biggl(\frac{1}{n}\biggr),
\end{split}
\end{equation}
which further implies the lower bound in Theorem~\ref{thm:mini_statistical}.

\section{Generalization}\label{sec:general}
In this section, we consider a generalization on the discrete database model and analyze the corresponding minimax distortion.

\subsection{Continuous Data Universe}
Consider databases with data universe $\datauniverse$ being an interval in the $\datauniversedim$ dimensional real coordinate space $\realspace{l}$. We assume that $l$ is a constant, so we present the case that $l=1$ and $\datauniverse=[0,1]$ for clarity. Consider the class of statistical queries with $L$-Lipschitz row functions, i.e., the query class
\begin{equation}
\mathcal{Q}^{\mathrm{S}}_L=\{q_{\varphi}\in\mathcal{Q}^{\mathrm{S}}\mid |\varphi_i(u)-\varphi_i(v)|\le L|u-v|,
\text{ for any }u,v\in[0,1]\text{ and any }i=1,2,\dots\}.
\end{equation}
Then the minimax distortion can be written as
\begin{equation*}
\mathfrak{D}_{\epsilon,L}^{\mathrm{S}}=\inf_{\mecmapping_{\mathcal{M}}\in\dpmapping}\sup_{q_{\varphi}\in\mathcal{Q}^{\mathrm{S}}_L,x\in\databaseuniverse}\expect_{Y\sim\mecmapping_{\mathcal{M}}(x)}\bigl[|\hat{q}_{\varphi}^*(Y)-q_{\varphi}(x)|^2\bigr].
\end{equation*}

We note that the lower bound in Theorem~\ref{thm:mini_statistical} still holds for continuous data universe since $\{0,1\}^n\subseteq[0,1]^n$. To obtain an upper bound, we consider the following approach for a database $x\in[0,1]^n$: first each row of $x$ is discretized into $k$ bits; then the mechanism $\mathcal{E}$ and the companion estimator $\hat{q}_{\varphi}$ are used for the discretized database. Denote the discretized database by $\hat{x}$. Then $\hat{x}\in\{0,\frac{1}{2^k},\dots,\frac{2^k-1}{2^k}\}^n$. By the discretization precision, $|x_i-\hat{x}_i|\le\frac{1}{2^k}$ for $i=1,2,\dots,n$. Thus for any $q_{\varphi}\in\mathcal{Q}^{\mathrm{S}}_L$,
\begin{align*}
|q_{\varphi}(x)-q_{\varphi}(\hat{x})|\le\frac{1}{\sum_{i=1}^{n}c_i}\sum_{i=1}^{n}|\varphi_i(x_i)-\varphi_i(\hat{x}_i)|\le\frac{L}{c2^k}.
\end{align*}
By Lemma~\ref{lem:qhat_distortion},
\begin{equation*}
\expect\bigl[|\hat{q}_{\varphi}(Y)-q_{\varphi}(\hat{x})|\bigr]\le\frac{2(b-a)\bigl(1+(2^k-1)e^{-\epsilon}\bigr)}{c(1-e^{-\epsilon})}\frac{1}{\sqrt{n}},
\end{equation*}
where we omit the subscript $Y\sim p_{\mathcal{E}(x)}$ of the expectation for conciseness. Then
\begin{align*}
&\mspace{24mu}\expect\bigl[|\hat{q}_{\varphi}(Y)-q_{\varphi}(x)|^2\bigr]\\
&\le\expect\bigl[\bigl(|q_{\varphi}(x)-q_{\varphi}(\hat{x})|+|\hat{q}_{\varphi}(Y)-q_{\varphi}(\hat{x})|\bigr)^2\bigr]\\
&\le|q_{\varphi}(x)-q_{\varphi}(\hat{x})|^2+|q_{\varphi}(x)-q_{\varphi}(\hat{x})|\cdot\expect[|\hat{q}_{\varphi}(Y)-q_{\varphi}(\hat{x})|]\\
&\mspace{24mu}+(\expect[|\hat{q}_{\varphi}(Y)-q_{\varphi}(\hat{x})|])^2\\
&\le\frac{L^2}{c^22^{2k}}+\frac{2(b-a)\bigl(1+(2^k-1)e^{-\epsilon}\bigr)L}{c^2(1-e^{-\epsilon})2^k}\frac{1}{\sqrt{n}}\\
&\mspace{24mu}+\frac{4(b-a)^2\bigl(1+(2^k-1)e^{-\epsilon}\bigr)^2}{c^2(1-e^{-\epsilon})^2}\frac{1}{n}.
\end{align*}
Let $2^{2k}=\sqrt{n}$. We obtain
\begin{equation*}
\expect\bigl[|\hat{q}_{\varphi}(Y)-q_{\varphi}(x)|^2\bigr]\le\biggl(\frac{L^2}{c^2}+\frac{4(b-a)^2e^{-2\epsilon}}{c^2(1-e^{-\epsilon})^2}\biggr)\frac{1}{\sqrt{n}}+o\biggl(\frac{1}{\sqrt{n}}\biggr),
\end{equation*}
which gives an upper bound on $\mathfrak{D}_{\epsilon,L}^{\mathrm{S}}$.
\begin{prop}
With continuous data universe $\datauniverse=[0,1]$, the minimax distortion for statistical queries with $L$-Lipschitz row functions satisfies the following bounds:
\begin{equation}
\frac{\bigl(1-\gausscdf(1)\bigr)^2}{ 2^{5}(1+e^{\epsilon})^3}\frac{1}{n}+o\biggl(\frac{1}{n}\biggr)\le \mathfrak{D}_{\epsilon,L}^{\mathrm{S}}
\le\biggl(\frac{L^2}{c^2}+\frac{4(b-a)^2e^{-2\epsilon}}{c^2(1-e^{-\epsilon})^2}\biggr)\frac{1}{\sqrt{n}}+o\biggl(\frac{1}{\sqrt{n}}\biggr).
\end{equation}
\end{prop}
\emph{Remark.} For a continuous data universe, the optimal estimator $\hat{q}_{\varphi}^*$ may not be attainable. In this case, we need to express the minimax distortion $\mathfrak{D}_{\epsilon,L}^{\mathrm{S}}$ in the same form as \eqref{eq:fullMinimax}. However, this does not change the arguments for the lower and upper bounds.

\section{Experimental Evaluation and Application}\label{sec:experiment}
In this section, we first evaluate the mechanism $\mathcal{E}$ in \eqref{eq:expMec} when companioned with the estimator $\hat{q}_{\varphi}^{\mathrm{u}}$ in \eqref{eq:qhatu} through experiments on a Netflix dataset \cite{Netflix} for statistical queries. During the experiments, we compare our approach with the MWEM algorithm (a combination of the Exponential Mechanism with the Multiplicative Weights update rules) \cite{HarLigMcS_12}. The main conclusion from the experimental results is that the proposed approach provides reasonable accuracy for all the tested queries, irrespective of the form of the queries or the number of the tested queries, which improves over the MWEM algorithm. The scaling behavior $\bigo(1/n)$ of the minimax distortion as the database size $n$ goes to infinity is also verified by the experimental results.

We next consider the application of differentially private cut function release for graphs and derive an upper bound on the minimax distortion for this application. We evaluate our approach through experiments on a Facebook dataset \cite{McALes_12}. The experimental results verify the theoretical upper bound and show that the proposed approach works well for this application.

\subsection{Evaluation for Statistical Queries}
In this subsection, we conduct experiments on the Netflix dataset for statistical queries. The Netflix dataset consists of movie ratings from users, with each rating on a scale from $1$ to $5$ (integral) stars. We treat each rating as a row and model the dataset as a database. To obtain databases with different sizes, we take subsets from the dataset.

The experimental evaluation in this subsection has three focuses: (1) the separation between statistical queries and linear queries, (2) distortion under varying query set size, and (3) scaling behavior of the distortion under varying database size.

\subsubsection{Statistical Queries vs. Linear Queries}
The class of statistical queries is much larger than the class of linear queries since a statistical query allows different row functions, whereas a linear query can only have identical row functions. For the private movie rating release application, it is possible to encounter queries that perform different functions on different rows, since different movies or users may belong to different groups and have different weights in a query. We call the number of distinct row functions in a statistical query the \emph{heterogeneity} of the query. Consider a statistical query $q_{\varphi}$ and the associated row function sequence $\varphi=(\varphi_1,\dots,\varphi_n)$. If the heterogeneity of $q_{\varphi}$ equals to $1$, then $\varphi_1=\dots=\varphi_n$, and thus $q_{\varphi}$ is a linear query. If the heterogeneity of $q_{\varphi}$ is greater than $1$, then not all the $\varphi_i$'s are equal. For example, during the experiments in this subsection, when the heterogeneity equal to $2$, the statistical query performs one row function for the first half of the rows, and performs another row function for the second half.

The mechanism $\mathcal{E}$ and the companion estimator $\hat{q}_{\varphi}^{\mathrm{u}}$ is designed for statistical queries. The upper bound \eqref{eq:qhatu_dis} on the distortion of the proposed approach holds for any statistical query, and thus holds for any heterogeneity. The MWEM algorithm is designed for linear queries. To evaluate the MWEM algorithm for statistical queries, we adapt it as follows. For each distinct row function in a statistical query, we treat the set of rows associated with this row function as a ``sub-database''. Restricted to this sub-database, the statistical query is a linear query, so we can run the MWEM algorithm on the sub-database to generate a synthetic sub-database. Then the answer to the statistical query is obtained by combining the answers at each sub-database. When there are multiple statistical queries in the query set, we need to divide the database into sub-databases such that restricted to a sub-database, any query in the query set is a linear query. In the experiment, we consider statistical queries with same row functions for ratings of the same movie.

We evaluate the proposed approach and the MWEM algorithm on a database of size $n=162,567$ from the Netflix dataset, consisting of ratings for $128$ movies. Each movie has roughly $1000\sim 2000$ ratings. Statistical queries are generated randomly in the following way. To specify a row function $\varphi_i$, the values $\varphi(1),\varphi(2),\dots,\varphi(5)$ are sufficient. We generate i.i.d. random variables $X_1,\dots,X_5$ with uniform distribution on $[0,1]$, and divide them by $\max_i X_i - \min_i X_i$ for normalization. Then these values are used to specify a row function. For a statistical query with heterogeneity $h$, we generate $h$ row functions independently, and assign each row function to rows corresponding to $1/h$ of the movies. During the experiments, we consider heterogeneity varying from $1$ to $128$. For each heterogeneity $h$, we generate a set of $200$ statistical queries with heterogeneity $h$ independently. We use the absolute-error distortion measure, i.e., $\querydis(s,t)=|s-t|$ for any $s,t\in\realnumber$, since both our approach and the MWEM algorithm have distortion upper bound under this distortion measure. We measure the worst-case distortion among the queries in the query set, and then take an average over $20$ independent runs. The differential privacy level is fixed to $\epsilon = 1$.

\begin{figure}[t]
\centering
\includegraphics[scale=\figurescaleparameter]{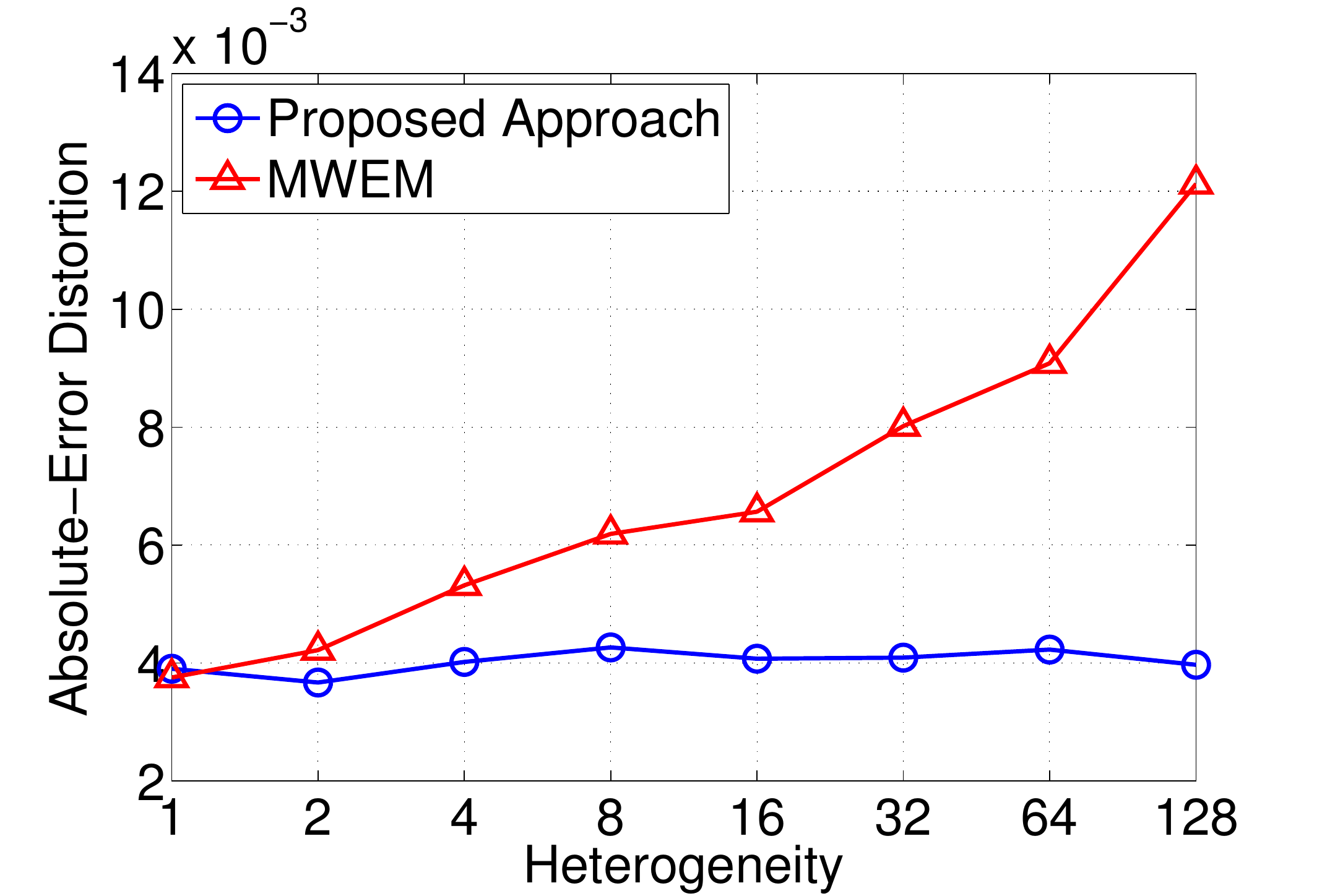}
\caption{Distortion under varying heterogeneity. The proposed approach is robust to heterogeneity, whereas the distortion of the MWEM algorithm grows as the heterogeneity increases.}
\label{fig:Heterogeneity}
\end{figure}
Figure~\ref{fig:Heterogeneity} compares our approach against the MWEM algorithm with varying heterogeneity. The figure shows that the proposed approach gives similar distortions irrespective of the heterogeneity, whereas under the MWEM algorithm, the distortion grows as the heterogeneity increases. This experimental result shows a separation between statistical queries and linear queries: approaches designed for linear queries cannot be directly applied to statistical queries without performance loss.

\subsubsection{Query Set Size--Independent Distortion}
Under most existing mechanisms \cite{BluLigRot_08,DwoNaoRei_09,HarLigMcS_12,GupRotUll_12,GabAriHsu_14} for synthetic database release, the accuracy guarantee becomes worse as the query set size increases. Under the MWEM algorithm, the worst-case distortion among the queries in a query set is $\bigo((\log(|\mathcal{Q}|))^{1/3})$, where $|\mathcal{Q}|$ is the query set size. Our approach does not restrict to a specific query set. The distortion upper bound in \eqref{eq:qhatu_dis} holds for all the statistical queries. Therefore, under our approach, the worst-case distortion among the queries in a query set will not grow as the query set size increases.

We evaluate the proposed approach and the MWEM algorithm on databases from the Netflix dataset. We randomly generate linear query sets with the size varying from $64$ to $1,048,576$, using the same method as in the previous experiments. We still use the absolute-error distortion measure. We measure the worst-case distortion among the queries in the query set and among $50$ databases, with database sizes roughly within $1000\sim 2000$. Then the worst-case distortion is averaged over $20$ independent runs. The differential privacy level is fixed to $\epsilon=1$.

\begin{figure}[t]
\centering
\includegraphics[scale=\figurescaleparameter]{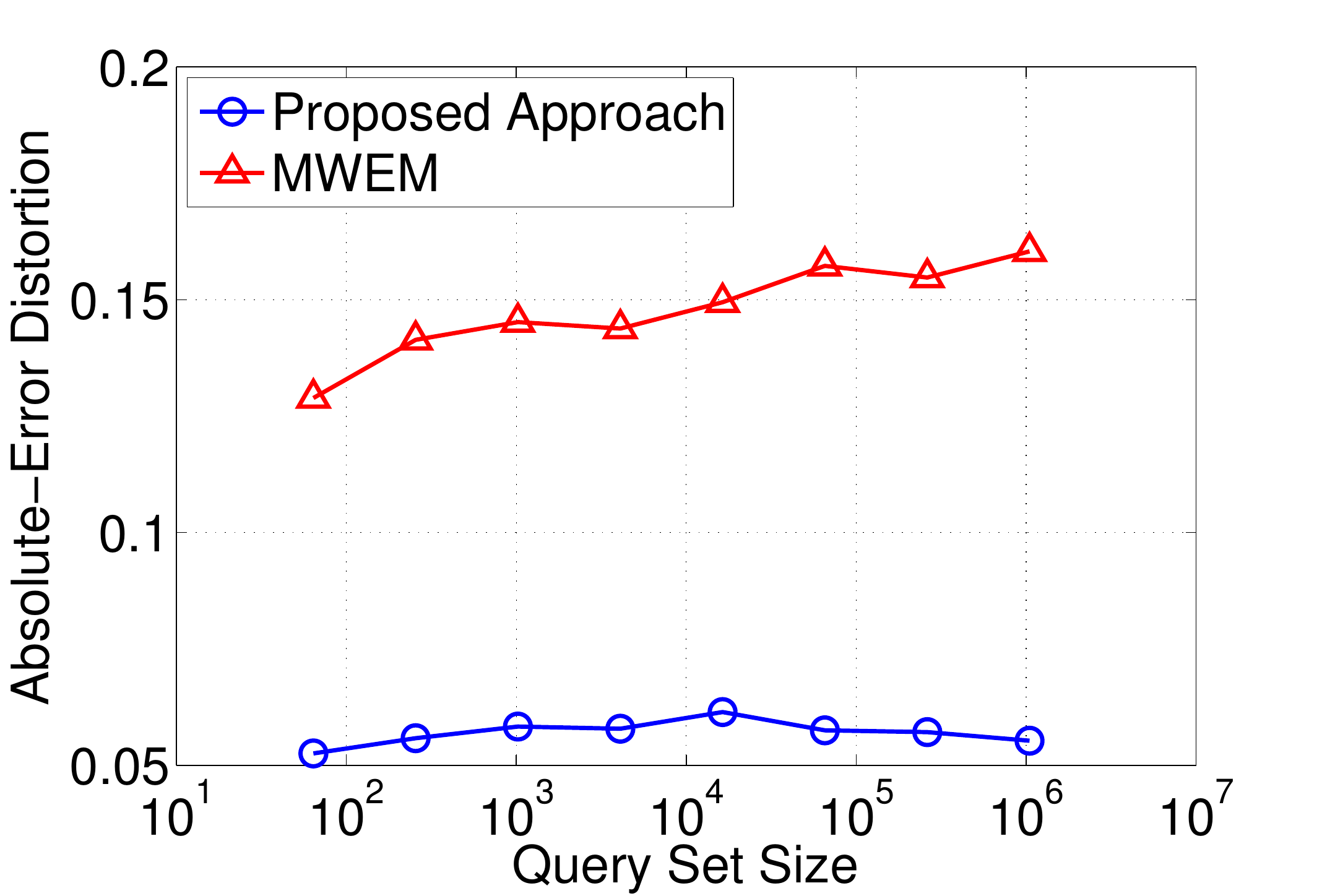}
\caption{Distortion under varying query set size. The worst-case distortion of the proposed approach does not depend on the query set size, whereas the distortion of the MWEM algorithm grows (slowly) as the query set size increases.}
\label{fig:QuerySetSize}
\end{figure}
Figure~\ref{fig:QuerySetSize} compares our approach against the MWEM algorithm with varying query set size. The figures shows that the proposed approach gives similar worst-case distortion for different query set sizes. However, for the MWEM algorithm, although very slowly, the worst-case distortion grows as the query set size increases. Therefore, to achieve certain accuracy, this growth indicates that the query set size must be smaller than a threshold. This experimental result verifies the dependence of the distortion on the query set size under the MWEM algorithm, and shows the advantage of our approach.

\subsubsection{Scaling Behavior}
\begin{figure}[t]
\centering
\includegraphics[scale=\figurescaleparameter]{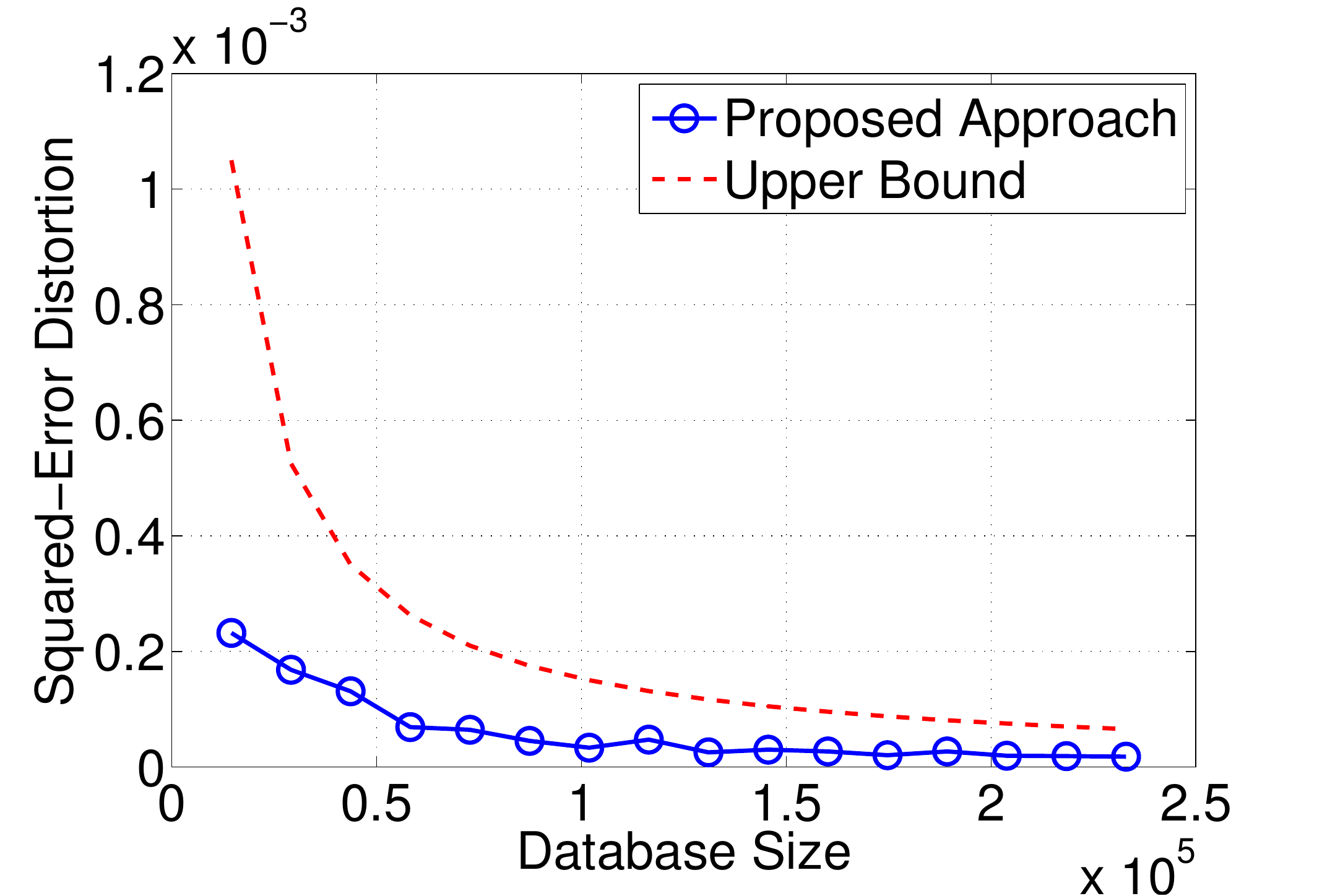}
\caption{Distortion under varying database size. In the asymptotic regime that the database size $n$ goes to infinity, the upper bound is $\Theta(1/n)$, so the distortion is $\bigo(1/n)$.}
\label{fig:DatabaseSize}
\end{figure}
Consider the asymptotic regime that the database size $n$ goes to infinity for given data universe dimension and differential privacy level. We have proved that the worst-case squared-error distortion of the mechanism $\mathcal{E}$ when companioned with the estimator $\hat{q}_{\varphi}^{\mathrm{u}}$ is $\bigo(1/n)$. To verify this theoretical upper bound, we evaluate the proposed approach on databases from the Netflix dataset. The sizes of the databases vary from $14,559$ to $232,944$. A linear query set of size $200$ is randomly generated in the same way as the previous experiments and used for all the databases. We use the squared-error distortion measure, i.e., $\rho(s,t)=(s-t)^2$ for any $s,t\in\realnumber$. We measure the worst-case distortion among the queries in the query set, and then take an average over $20$ independent runs. The differential privacy level is fixed to $\epsilon=1$. Figure~\ref{fig:DatabaseSize} compares the distortion under the proposed approach with the upper bound in \eqref{eq:qhatu_dis}, which verifies the asymptotic order $\bigo(1/n)$ of the distortion.

\subsection{Differentially Private Cut Function Release for Graphs}\label{sec:cut}
Consider the scenario that the given database is a graph, where the presence of individual edges is sensitive information. Such a graph can represent the online social connections between individuals. To release useful information for graph analysis, a well studied approach is to privately release the cut function of the graph \cite{BloBluDat_12,GupHarRot_11,GupRotUll_12}.

Let the graph be $G=(V,E)$ and $\powerset{V}$ denote the power set of $V$. Then the cut function $f_G\colon \powerset{V}\times\powerset{V}\rightarrow[|E|]$ associated with this graph is defined by
\begin{equation}
f_G(S,T) = |\{(i,j)\in E\mid i\in S, j\in T\}|,
\end{equation}
which is the number of edges crossing the $S,T$-cut for any disjoint $S,T\subseteq V$.

We use a database $x$ to represent the graph $G$. Since differential privacy needs to be preserved for edges, each row of $x$ corresponds to a vertex pair $(i,j)\in V\times V$, where $x_{i,j}=1$ if $(i,j)\in E$, and $x_{i,j}=0$ otherwise. Here we use $(i,j)$ to index each row of $x$. Thus the data universe is $\{0,1\}$ with dimension $l=1$ and the database size $n=|V|^2$. Two databases $x,x'$ are neighbors if there exists exactly one vertex pair $(i,j)$ such that $x_{i,j}\neq x'_{i,j}$.

For any disjoint $S,T\subseteq V$, we write $f_G(S,T)$ as a function $q_{S,T}$ of $x$ and call it a \emph{cut query}. Consider the absolute-error distortion measure $\rho(s,t)=|s-t|$ for any $s,t\in\realnumber$. Then the minimax distortion for $\epsilon$-differentially private cut function release can be written as
\begin{equation*}
\minimax^{\mathrm{C}}=\inf_{\mecmapping_{\mathcal{M}}\in\dpmapping}\sup_{\substack{x\in\{0,1\}^n\\S,T\subseteq V,S\cap T=\emptyset}}\expect_{Y\sim\mecmapping_{\mathcal{M}}(x)}\bigl[|\hat{q}_{S,T}^*(Y)-q_{S,T}(x)|\bigr].
\end{equation*}

Consider the statistical query defined in Definition~\ref{def:stat_query}. Then a cut query $q_{S,T}$ can be viewed as an unnormalized statistical query over the subset $S\times T\subseteq V\times V$ of all the rows. The row function is $\varphi_{i,j}(x_{i,j})=x_{i,j}$ since
\begin{equation}
q_{S,T}(x) =\sum_{(i,j)\in S\times T}x_{i,j}.
\end{equation}
Consider the mechanism $\mathcal{E}$ and estimator $\hat{q}_{S,T}\colon\{0,1\}^n \rightarrow\realnumber$ defined by
\begin{equation}
\hat{q}_{S,T}(y)=\frac{1+e^{-\epsilon}}{1-e^{-\epsilon}}q_{S,T}(y)-\frac{e^{-\epsilon}}{1-e^{-\epsilon}}|S||T|,
\end{equation}
which is an adapted version of the estimator $\hat{q}^{\mathrm{u}}_{\varphi}$ defined in \eqref{eq:qhatu} for the query $q_{S,T}$. Let $Y$ denote the released synthetic database $\mathcal{E}(x)$. By similar analysis as in the proof of Lemma~\ref{lem:qhatu}, the distortion is bounded as
\begin{equation}
\expect_{Y\sim\mecmapping_{\mathcal{E}}(x)}\bigl[|\hat{q}_{S,T}(Y)-q_{S,T}(x)|\bigr]\le\frac{1+e^{-\epsilon}}{1-e^{-\epsilon}}\sqrt{|S||T|}.\label{eq:cut_dis}
\end{equation}
For any $S,T\subseteq V$, $|S||T|\le |V|^2$. Therefore the minimax distortion is upper bounded as
\begin{equation}
\minimax^{\mathrm{C}}\le\frac{1+e^{-\epsilon}}{1-e^{-\epsilon}}|V|.
\end{equation}

\subsubsection{Evaluation on the Facebook Dataset}
We evaluate the proposed approach on databases from the Facebook dataset for the application of cut query release. The Facebook dataset is a graph. Each vertex in the graph represents a user, and an edge between two vertices indicates that they are friends.

\begin{figure}[t]
\centering
\includegraphics[scale=\figurescaleparameter]{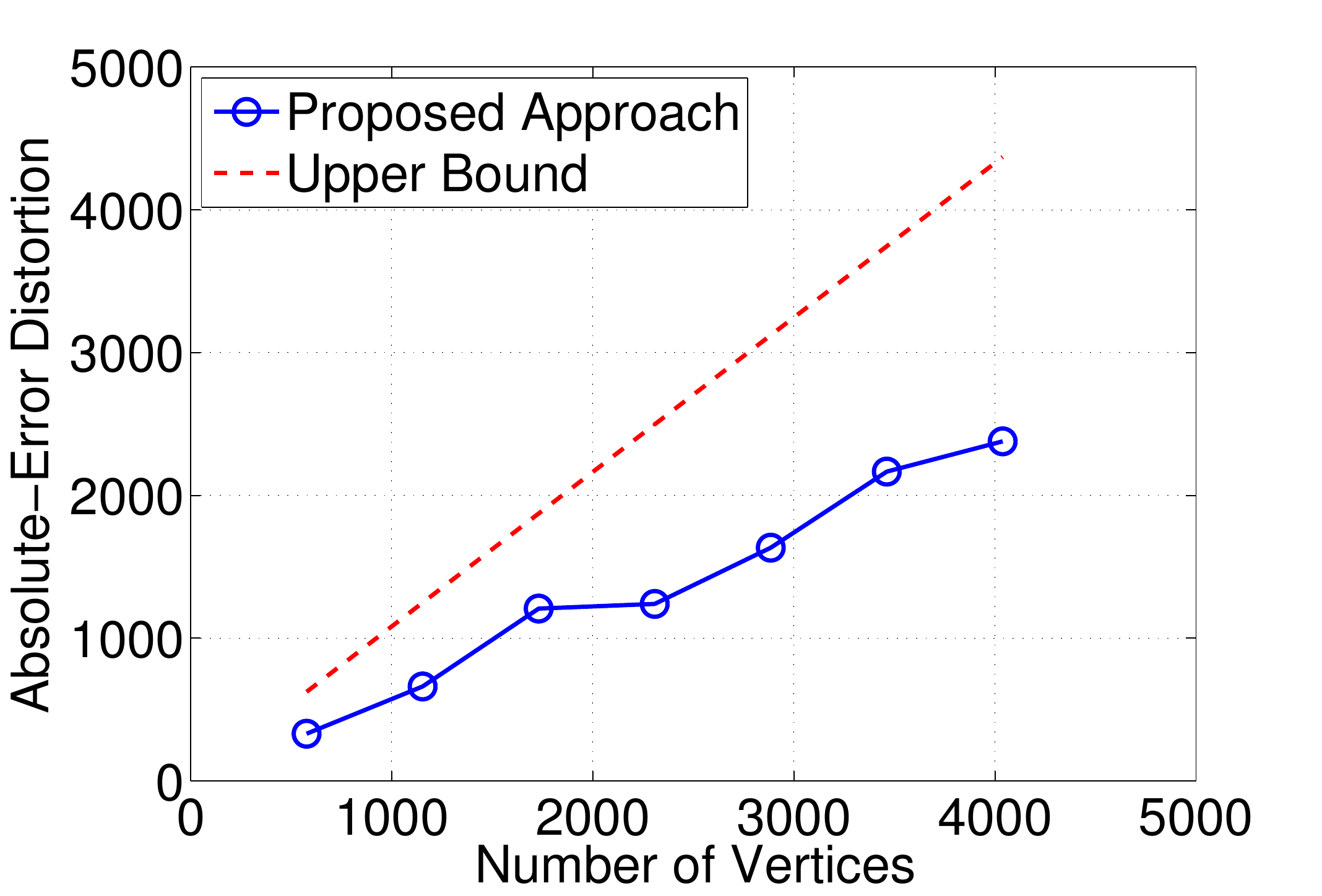}
\caption{Distortion of cut queries under varying number of nodes in the graph. In the asymptotic regime that the number of nodes $|V|$ goes to infinity, the upper bound is $\Theta(|V|)$, so the distortion is $\bigo(|V|)$.}
\label{fig:CutQueries}
\end{figure}

Consider the asymptotic regime that the number of vertices $|V|$ goes to infinity. We have proved that the absolute-error distortion for any cut query is $\bigo(|V|)$. To verify this theoretical upper bound, we apply our approach on subgraphs of the graph given by the Facebook dataset. The graph consists of $4039$ vertices and $88,234$ edges. The number of vertices in the considered subgraphs vary from $577$ to $4039$. For each subgraph, cut queries are generated randomly in the following way. Half of the vertices are uniformly sampled and this vertex set is denoted by $S$. Then $S$ and $V-S$ specify a cut query. This choice of cut queries results in the largest upper bound on the distortion as shown in \eqref{eq:cut_dis}. We generate a cut query set consisting of $100$ cut queries independently. We measure the worst-case absolute-error distortion among the cut queries in the query set, and then take an average over $10$ independent runs. The differential privacy level is fixed to $\epsilon=1$. Figure~\ref{fig:CutQueries} compares the distortion under the proposed approach with the upper bound in \eqref{eq:cut_dis}, which verifies the asymptotic order $\bigo(|V|)$ of the distortion. The worst-case relative distortion in Table~\ref{tab:CutQueries} shows that the accuracy is reasonable for cut queries.
\begin{table}
\small
\centering
\begin{tabular}{cccccccc}
\toprule
$|V|$ & 577 & 1154 & 1731 & 2308 & 2885 & 3462 & 4039 \\
\midrule
Error & 10.4\% & 11.7\% & 8.7\% & 5.3\% & 4.7\% & 5.3\% & 5.4\% \\
\bottomrule
\end{tabular}
\caption{Relative error for cut queries.}
\label{tab:CutQueries}
\end{table}

\section{Conclusion and Future Work}\label{sec:conclusion}
In this paper, we developed a minimax approach for differentially private query release, where query-set independent differentially private synthetic database releasing mechanisms are devised and the companion estimators are designed to provide accurate answers for all queries in a general query class. For the general class of statistical queries, we proved that with the squared-error distortion measure, the minimax distortion $\minimax^{\mathrm{S}}$ is $\bigo(1/n)$ by deriving asymptotically tight upper and lower bounds in the regime that the database size $n$ goes to infinity. The upper bound was achieved by a differentially private synthetic database releasing mechanism $\mathcal{E}$ and the companion estimators, which indicates that it is feasible to use query-set independent differentially private synthetic database releasing mechanisms while providing accurate answers for all the statistical queries in large databases.

In this work, we have focused on the minimax distortion in the asymptotic regime that database size $n$ grows. It is also of great interest to quantify the scaling laws of the minimax distortion in the joint asymptotic regime in terms of database size $n$, data universe dimension $l$ and the differential privacy level $\epsilon$. We are currently investigating this issue and aim at designing better differentially private synthetic database releasing mechanisms for large data universe dimension $l$ and finding tighter lower bounds in terms of $\epsilon$.

\bibliographystyle{IEEEtran}
\bibliography{differential-privacy}

\appendices
\section{Proof of Lemma~2}\label{app:lem_qhatu}
\begin{proof}
We drop the subscript $Y\sim\mu_{\mathcal{E}(x)}$ from expectations for conciseness during the proof. We first prove that the estimator $\hat{q}_{\varphi}^{\mathrm{u}}$ is unbiased. Recall that $\{Y_i,i\in[n]\}$ follow the \pmf s in \eqref{eq:pmf_row}. Then
\begin{align*}
\expect[q_{\varphi}(Y)]&=\frac{1}{\sum_{i=1}^{n}c_i}\sum_{i=1}^{n}\expect[\varphi_i(Y_i)]\\
&=\frac{1}{\sum_{i=1}^{n}c_i}\sum_{i=1}^{n}\biggl(\frac{1}{\denom}\varphi_i(x_i)+\frac{e^{-\epsilon}}{\denom}\sum_{\substack{\entry\in\datauniverse\colon\\\mspace{-6mu}\entry\neq x_i}}\varphi_i(\entry)\biggr)\\
&=\frac{1-e^{-\epsilon}}{\denom}\frac{1}{\sum_{i=1}^{n}c_i}\sum_{i=1}^{n}\varphi_i(x_i)\\
&\mspace{22mu}+\frac{e^{-\epsilon}}{\denom}\frac{1}{\sum_{i=1}^{n}c_i}\sum_{i=1}^{n}\sum_{\entry\in\datauniverse}\varphi_i(\entry)\\
&=\frac{1-e^{-\epsilon}}{\denom}q_{\varphi}(x)+\frac{e^{-\epsilon}}{g(\epsilon)}C_{\varphi}.
\end{align*}
Therefore
\begin{equation*}
\expect[\hat{q}_{\varphi}^{\mathrm{u}}(Y)]=\expect\biggl[\frac{\denom}{1-e^{-\epsilon}}q_{\varphi}(Y)-\frac{e^{-\epsilon}}{1-e^{-\epsilon}}C_{\varphi}\biggr]=q_{\varphi}(x).
\end{equation*}

Next we prove the upper bound on the distortion of $\hat{q}_{\varphi}^{\mathrm{u}}$. For any $x\in\databaseuniverse$,
\begin{align*}
&\mspace{24mu}\hat{q}_{\varphi}^{\mathrm{u}}(Y)-q_{\varphi}(x)\\
&=\frac{\denom}{1-e^{-\epsilon}}\frac{1}{\sum_{i=1}^{n}c_i}\\
&\mspace{24mu}\cdot\sum_{i=1}^{n}\biggl(\varphi_i(Y_i)-\frac{1-e^{-\epsilon}}{\denom}\varphi_i(x_i)-\frac{e^{-\epsilon}}{\denom}\sum_{\entry\in\datauniverse}\varphi_i(\entry)\biggr).
\end{align*}
For any $i\in[n]$, let
\begin{equation*}
Z_i=\varphi_i(Y_i)-\frac{1-e^{-\epsilon}}{\denom}\varphi_i(x_i)-\frac{e^{-\epsilon}}{\denom}\sum_{\entry\in\datauniverse}\varphi_i(\entry).
\end{equation*}
Then for any $i\in[n]$, $\expect[Z_i]=0$. Recall that for any $\entry\in\datauniverse$, $a\le\varphi_i(\entry)\le b$, so $|Z_i|\le b-a$. Since $Y_1,\dots,Y_n$ are independent, $Z_1,\dots,Z_n$ are independent. Let $\overline{Z}=\frac{1}{n}\sum_{i=1}^{n}Z_i$. Then
\begin{align*}
&\mspace{24mu}\expect\bigl[|\hat{q}_{\varphi}^{\mathrm{u}}(Y)-q_{\varphi}(x)|^2\bigr]\\
&=\biggl(\frac{\denom}{1-e^{-\epsilon}}\frac{n}{\sum_{i=1}^{n}c_i}\biggr)^{2}\cdot\expect\bigl[\bigl|\overline{Z}\bigr|^2\bigr]\\
&=\biggl(\frac{\denom}{1-e^{-\epsilon}}\frac{n}{\sum_{i=1}^{n}c_i}\biggr)^{2}\cdot\biggl(\frac{1}{n^2}\sum_{i=1}^{n}\expect\bigl[|Z_i|^2\bigr]\biggr)\\
&\le\biggl(\frac{\denom}{1-e^{-\epsilon}}\biggr)^{2}\frac{1}{c^2}\frac{(b-a)^2}{n}.
\end{align*}
Therefore
\begin{equation*}
\sup_{x\in\databaseuniverse}\expect\bigl[|\hat{q}_{\varphi}^{\mathrm{u}}(Y)-q_{\varphi}(x)|^2\bigr]\le\frac{(b-a)^2\bigl(1+(2^l-1)e^{-\epsilon}\bigr)^2}{c^2(1-e^{-\epsilon})^2}\frac{1}{n}.
\end{equation*}
\end{proof}

\section{Proof of Lemma~3}\label{app:lem_qhat}
\begin{proof}
For any $x,y\in\databaseuniverse$, since $q_{\varphi}(x)\in q_{\varphi}(\databaseuniverse)$, by the definition of the estimator $\hat{q}_{\varphi}$ in \eqref{eq:qhat},
\begin{equation*}
|\hat{q}_{\varphi}^{\mathrm{u}}(y)-\hat{q}_{\varphi}(y)|\le|\hat{q}_{\varphi}^{\mathrm{u}}(y)-q_{\varphi}(x)|.
\end{equation*}
Therefore
\begin{align*}
|\hat{q}_{\varphi}(y)-q_{\varphi}(x)|&\le|\hat{q}_{\varphi}(y)-\hat{q}_{\varphi}^{\mathrm{u}}(y)|+|\hat{q}_{\varphi}^{\mathrm{u}}(y)-q_{\varphi}(x)|\\
&\le 2|\hat{q}_{\varphi}^{\mathrm{u}}(y)-q_{\varphi}(x)|,
\end{align*}
and
\begin{equation*}
\expect\bigl[|\hat{q}_{\varphi}(Y)-q_{\varphi}(x)|^2\bigr]\le 4\expect\bigl[|\hat{q}_{\varphi}^{\mathrm{u}}(Y)-q_{\varphi}(x)|^2\bigr].
\end{equation*}
Then combining with \eqref{eq:qhatu_dis} yields the upper bound.
\end{proof}

\section{Proof of Lemma~\ref{lem:lower}}\label{app:lem_lower}
\begin{proof}
By Jensen's inequality,
\begin{align}
&\mspace{25mu}\expect\bigl[|\expect[d(X,\Queryvar)\mid Y,\Queryvar]-d(X,\Queryvar)|^2\bigr]\nonumber\\
&\ge\bigl(\expect\bigl[|\expect[d(X,\Queryvar)\mid Y,\Queryvar]-d(X,\Queryvar)|\bigr]\bigr)^2.\label{eq:jensen}
\end{align}
Let $\ReconX$ be a random variable satisfying the following conditions: $\ReconX$ is independent of $\Queryvar$; $\ReconX$ is independent of $X$ given $Y$; given $Y$, $\ReconX$ and $X$ are identically distributed, i.e., $p_{\ReconX\mid Y}(x\mid y)=p_{X\mid Y}(x\mid y)$ for any $x,y\in\databaseuniverse$ with $p_Y(y)\neq 0$. Due to the independence between $\Queryvar$ and $(X,Y,\ReconX)$, we also have $p_{\ReconX\mid Y,\Queryvar}(x\mid y,\queryvar)=p_{X\mid Y,\Queryvar}(x\mid y,\queryvar)$ for any $x,y,\queryvar\in\databaseuniverse$ with $p_Y(y)\neq 0$. By this construction, for any $y,\queryvar\in\databaseuniverse$ with $p_Y(y)\neq 0$,
\begin{align*}
\expect[d(X,\Queryvar)\mid Y=y,\Queryvar=\queryvar]=\expect[d(\ReconX,\Queryvar)\mid Y=y,\Queryvar=\queryvar],
\end{align*}
and
\begin{align*}
&\mspace{25mu}\expect\bigl[|\expect[d(X,\Queryvar)\mid Y,\Queryvar]-d(X,\Queryvar)|\bigm| Y=y,\Queryvar=\queryvar\bigr]\\
&=\expect\bigl[|\expect[d(\ReconX,\Queryvar)\mid Y,\Queryvar]-d(\ReconX,\Queryvar)|\bigm| Y=y,\Queryvar=\queryvar\bigr],
\end{align*}
which further lead to
\begin{align*}
&\mspace{25mu}\expect\bigl[|\expect[d(X,\Queryvar)\mid Y,\Queryvar]-d(X,\Queryvar)|\bigr]\\
&=\expect\Bigl[\expect\bigl[|\expect[d(X,\Queryvar)\mid Y,\Queryvar]-d(X,\Queryvar)|\bigm| Y,\Queryvar\bigr]\Bigr]\\
&=\expect\Bigl[\expect\bigl[|\expect[d(\ReconX,\Queryvar)\mid Y,\Queryvar]-d(\ReconX,\Queryvar)|\bigm| Y,\Queryvar\bigr]\Bigr]\\
&=\expect\bigl[|\expect[d(\ReconX,\Queryvar)\mid Y,\Queryvar]-d(\ReconX,\Queryvar)|\bigr].
\end{align*}
Therefore
\begin{align*}
&\mspace{25mu}2\expect\bigl[|\expect[d(X,\Queryvar)\mid Y,\Queryvar]-d(X,\Queryvar)|\bigr]\\
&=\expect\bigl[|\expect[d(X,\Queryvar)\mid Y,\Queryvar]-d(X,\Queryvar)|\\
&\mspace{42mu}+|\expect[d(\ReconX,\Queryvar)\mid Y,\Queryvar]-d(\ReconX,\Queryvar)|\bigr]\\
&\ge\expect\bigl[|d(X,Z)-d(\ReconX,Z)\\
&\mspace{42mu}+\expect[d(\ReconX,\Queryvar)\mid Y,\Queryvar]-\expect[d(X,\Queryvar)\mid Y,\Queryvar]|\bigr]\\
&=\expect\bigl[|d(X,Z)-d(\ReconX,Z)|\bigr].
\end{align*}
Combing this with \eqref{eq:jensen} gives
\begin{align}
&\mspace{25mu}\expect\bigl[|\expect[d(X,\Queryvar)\mid Y,\Queryvar]-d(X,\Queryvar)|^2\bigr]\nonumber\\
&\ge\frac{1}{4}\bigl(\expect\bigl[|d(X,Z)-d(\ReconX,Z)|\bigr]\bigr)^2.\label{eq:quarter}
\end{align}
Then it suffices to derive a lower bound on $\expect\bigl[|d(X,Z)-d(\ReconX,Z)|\bigr]$.

Notice that the conditional \pmf\ $p_{\ReconX\mid X}$ is $\epsilon$-differentially private since for any neighboring $x,x'\in\databaseuniverse$ and any $\reconX\in\databaseuniverse$,
\begin{align}
p_{\ReconX\mid X}(\reconX\mid x)&=\sum_{y\in\databaseuniverse}p_{\ReconX\mid Y,X}(\reconX\mid y,x)p_{Y\mid X}(y\mid x)\\
&=\sum_{y\in\databaseuniverse}p_{\ReconX\mid Y,X}(\reconX\mid y,x')p_{Y\mid X}(y\mid x)\label{eq:con-ind}\\
&\le\sum_{y\in\databaseuniverse}p_{\ReconX\mid Y,X}(\reconX\mid y,x')\cdot e^{\epsilon}p_{Y\mid X}(y\mid x')\label{eq:diff-private}\\
&=e^{\epsilon}p_{\ReconX\mid X}(\reconX\mid x'),
\end{align}
where \eqref{eq:con-ind} follows from the conditional independence between $\ReconX$ and $X$ given $Y$, and \eqref{eq:diff-private} holds because $p_{Y\mid X}$ is $\epsilon$-differentially private. Then by Theorem~1 in \cite{WanYinZha_14_3} (for our case, the $\epsilon_X$ in that theorem is $0$),
\begin{equation*}
\expect[d(X,\ReconX)]\ge\frac{n}{1+\frac{e^{\epsilon}}{2^l-1}}.
\end{equation*}
Let $\gamma=\frac{1}{2(1+\frac{e^{\epsilon}}{2^l-1})}$ and $s=\gamma n$. Since
\begin{align*}
\expect[d(X,\ReconX)]&\le s\Pr\{d(X,\ReconX)<s\}+n\Pr\{d(X,\ReconX)\ge s\}\\
&\le s+n\Pr\{d(X,\ReconX)\ge s\},
\end{align*}
we have
\begin{align*}
\Pr\{d(X,\ReconX)\ge s\}&\ge\frac{1}{n}(\expect[d(X,\ReconX)]-s)\\
&\ge\frac{1}{n}\Biggl(\frac{n}{1+\frac{e^{\epsilon}}{2^l-1}}-\frac{n}{2\bigl(1+\frac{e^{\epsilon}}{2^l-1}\bigr)}\Biggr)\\
&=\gamma,
\end{align*}
i.e.,
\begin{equation}\label{eq:prob-lower}
\Pr\{d(X,\ReconX)\ge \gamma n\}\ge\gamma.
\end{equation}
We will consider those $x,\reconX\in\databaseuniverse$ with $d(x,\reconX)\ge\gamma n$ to obtain a lower bound on $\expect[|d(X,\Queryvar)-d(\ReconX,\Queryvar)|]$.

Utilizing conditional expectation gives
\begin{align}
&\mspace{25mu}\expect[|d(X,\Queryvar)-d(\ReconX,\Queryvar)|]\nonumber\\
&=\expect\bigl[\expect[|d(X,\Queryvar)-d(\ReconX,\Queryvar)|\mid X,\ReconX]\bigr]\nonumber\\
&\ge\mspace{-12mu}\sum_{\substack{x,\reconX:\\d(x,\reconX)\ge \gamma n}}\mspace{-18mu}\expect[|d(X,\Queryvar)-d(\ReconX,\Queryvar)|\mid X=x,\ReconX=\reconX]p_{X,\ReconX}(x,\reconX).\label{eq:cond-exp}
\end{align}
Consider any $x,\reconX\in\databaseuniverse$ with $d(x,\reconX)\ge\gamma n$ and $p_{X,\ReconX}(x,\reconX)\neq 0$. Since $\Queryvar$ is independent of $(X,\ReconX)$,
\begin{align}
&\mspace{25mu}\expect[|d(X,\Queryvar)-d(\ReconX,\Queryvar)|\mid X=x,\ReconX=\reconX]\nonumber\\
&=\expect[|d(x,\Queryvar)-d(\reconX,\Queryvar)|].\label{eq:query-exp}
\end{align}
Let
\begin{equation}
\diff(x,\reconX)=\{i\in[n]\mid x_i\neq \reconX_i\}.
\end{equation}
Then $|\diff(x,\reconX)|\ge\gamma n$, and
\begin{align*}
|d(x,\Queryvar)-d(\reconX,\Queryvar)|&=\biggl|\sum_{i=1}^{n}\bigl(\hamentry(x_i,\Queryvar_i)-\hamentry(\reconX_i,\Queryvar_i)\bigr)\biggr|\\
&=\biggl|\sum_{i\in\diff(x,\reconX)}\bigl(\hamentry(x_i,\Queryvar_i)-\hamentry(\reconX_i,\Queryvar_i)\bigr)\biggr|.
\end{align*}
Let
\begin{equation}
U_i=\hamentry(x_i,\Queryvar_i)-\hamentry(\reconX_i,\Queryvar_i).
\end{equation}
Since $\Queryvar$ is uniformly distributed over $\databaseuniverse$, the rows $\Queryvar_1,\Queryvar_2,\dots,\Queryvar_n$ are i.i.d. with \pmf\ $p_{\Queryvar_i}(\queryvar_i)=\frac{1}{2^l}$ for any $\queryvar_i\in\datauniverse$. For any $i\in\diff(x,\reconX)$,
\begin{equation}
U_i=
\begin{cases}
1 & \text{if $\Queryvar_i=\reconX_i$,}\\
-1 & \text{if $\Queryvar_i=x_i$,}\\
0 & \text{otherwise.}
\end{cases}
\end{equation}
Therefore $\{U_i,i\in\diff(x,\reconX)\}$ are i.i.d. with \pmf\
\begin{equation}
p_{U_i}(u_i)=
\begin{cases}
\frac{1}{2^l} & u_i=1,\\
\frac{1}{2^l} & u_i=-1,\\
1-\frac{1}{2^{l-1}} & u_i=0.
\end{cases}
\end{equation}
Then $\expect[U_i]=0$. Denote
\begin{equation}
\sigma^2=\expect\bigl[|U_i|^2\bigr]=\frac{1}{2^{l-1}},\quad\rho=\expect\bigl[|U_i|^3\bigr]=\frac{1}{2^{l-1}}.
\end{equation}
By the Berry--Esseen theorem \cite[Theorem~7.4.1]{Chu_00}, there exists a universal constant $C$ such that for any $t$,
\begin{align*}
&\mspace{23mu}\Pr\Biggl\{\frac{1}{\sigma \sqrt{|\diff(x,\reconX)|}}\sum_{i\in\diff(x,\reconX)}U_i>\frac{t}{\sigma\sqrt{\gamma}}
\Biggr\}\\
&\ge 1-\gausscdf\Bigl(\frac{t}{\sigma\sqrt{\gamma}}\Bigr)-\frac{C\rho}{\sigma^3\sqrt{|\diff(x,\reconX)|}}\\
&\ge 1-\gausscdf\Bigl(\frac{t}{\sigma\sqrt{\gamma}}\Bigr)-\frac{C\rho}{\sigma^3\sqrt{\gamma n}},
\end{align*}
where the second inequality follows from $|\diff(x,\reconX)|\ge\gamma n$. Therefore
\begin{align*}
&\mspace{24mu}\Pr\bigl\{|d(x,\Queryvar)-d(\reconX,\Queryvar)|> t\sqrt{n}\bigr\}\\
&=\Pr\Biggl\{\frac{1}{\sigma \sqrt{|\diff(x,\reconX)|}}\sum_{i\in\diff(x,\reconX)}U_i>\frac{t\sqrt{n}}{\sigma\sqrt{|\diff(x,\reconX)|}}\Biggr\}\\
&\ge\Pr\Biggl\{\frac{1}{\sigma \sqrt{|\diff(x,\reconX)|}}\sum_{i\in\diff(x,\reconX)}U_i>\frac{t\sqrt{n}}{\sigma\sqrt{\gamma n}}\Biggr\}\\
&\ge 1-\gausscdf\Bigl(\frac{t}{\sigma\sqrt{\gamma}}\Bigr)-\frac{C\rho}{\sigma^3\sqrt{\gamma n}}.
\end{align*}
Let $t=\sigma\sqrt{\gamma}$, then
\begin{align*}
\Pr\bigl\{|d(x,\Queryvar)-d(\reconX,\Queryvar)| > \sigma\sqrt{\gamma n}\bigr\}\ge 1-\gausscdf(1)-\frac{C\rho}{\sigma^3\sqrt{\gamma n}},
\end{align*}
and further
\begin{align}
&\mspace{25mu}\expect[|d(x,\Queryvar)-d(\reconX,\Queryvar)|]\nonumber\\
&\ge\sigma\sqrt{\gamma n}\cdot\Pr\bigl\{|d(x,\Queryvar)-d(\reconX,\Queryvar)|>\sigma\sqrt{\gamma n}\bigr\}\nonumber\\
&\ge\bigl(1-\gausscdf(1)\bigr)\sigma\sqrt{\gamma n}-\frac{C\rho}{\sigma^3}.
\end{align}
Inserting this lower bound back to \eqref{eq:query-exp}, \eqref{eq:cond-exp} and combining the lower bound \eqref{eq:prob-lower} yield
\begin{align*}
&\mspace{25mu}\expect[|d(X,\Queryvar)-d(\ReconX,\Queryvar)|]\\
&\ge\mspace{-12mu}\sum_{\substack{x,\reconX:\\ d(x,\reconX)\ge\gamma n}}\mspace{-18mu}\biggl(\bigl(1-\gausscdf(1)\bigr)\sigma\sqrt{\gamma n}-\frac{C\rho}{\sigma^3}\biggr)p_{X,\ReconX}(x,\reconX)\\
&=\biggl(\bigl(1-\gausscdf(1)\bigr)\sigma\sqrt{\gamma n}-\frac{C\rho}{\sigma^3}\biggr)\Pr\{d(X,\ReconX)\ge\gamma n\}\\
&\ge\bigl(1-\gausscdf(1)\bigr)\sigma\gamma^{\frac{3}{2}}\sqrt{n}-\frac{C\rho\gamma}{\sigma^3}.
\end{align*}
Therefore, by \eqref{eq:quarter},
\begin{align*}
&\mspace{22mu}\expect\bigl[|\expect[d(X,\Queryvar)\mid Y,\Queryvar]-d(X,\Queryvar)|^2\bigr]\\
&\ge\frac{1}{4}\biggl(\bigl(1-\gausscdf(1)\bigr)\sigma\gamma^{\frac{3}{2}}\sqrt{n}-\frac{C\rho\gamma}{\sigma^3}\biggr)^2,
\end{align*}
which completes the proof.
\end{proof}
\end{document}